\documentclass[aps,prb,floatfix,reprint,amsmath,amssymb,superscriptaddress]{revtex4-2}
\usepackage{amsfonts}
\usepackage{mathrsfs}
\usepackage{amsmath}
\usepackage{color}
\usepackage{natbib}
\usepackage{graphicx}
\usepackage{bm}
\usepackage{amssymb}
\usepackage{xspace}
\usepackage{epstopdf}
\usepackage{dcolumn}
\usepackage{longtable}
\usepackage{multirow}
\usepackage{float}

\usepackage[colorlinks=true, letterpaper=true, pdfstartview=FitV, linkcolor=blue, citecolor=blue, urlcolor=blue]{hyperref}

\makeatother

\begin{document}

\title{Unusually high phonon thermal conductivity in the Weyl semimetal TaP: A comparative study with TaAs}

\author{Xianyong Ding}
\altaffiliation{X. Y. D. and X. J. contributed equally to this work.}
\affiliation{College of Physics and Optoelectronic Engineering, Chongqing Normal University, Chongqing 401331, China}

\author{Xin Jin}
\altaffiliation{X. Y. D. and X. J. contributed equally to this work.}
\affiliation{College of Physics and Optoelectronic Engineering, Chongqing Normal University, Chongqing 401331, China}

\author{Dengfeng Li}
\affiliation{School of Science, Chongqing University of Posts and Telecommunications, Chongqing, 400065, China}

\author{Jing Fan}
\affiliation{Center for Computational Science and Engineering, Southern University of Science and Technology, Shenzhen 518055, P. R. China}

\author{Peng Yu}
\affiliation{College of Physics and Electronic Engineering, Chongqing Normal University, Chongqing 401331, China}

\author{Xiaoyuan Zhou}
\affiliation{College of Physics and Center of Quantum Materials and Devices, Chongqing University, Chongqing 401331, China}

\author{Xiaolong Yang}
\email{yangxl@cqu.edu.cn}
\affiliation{College of Physics and Center of Quantum Materials and Devices, Chongqing University, Chongqing 401331, China}
\affiliation{Chongqing Key Laboratory for Strongly Coupled Physics, Chongqing University, Chongqing 401331, China}

\author{Rui Wang}
\email{rcwang@cqu.edu.cn}
\affiliation{College of Physics and Center of Quantum Materials and Devices, Chongqing University, Chongqing 401331, China}
\affiliation{Chongqing Key Laboratory for Strongly Coupled Physics, Chongqing University, Chongqing 401331, China}

\begin{abstract}
In many metals, thermal transport is often dominated by electrons, although the lattice contribution can remain appreciable depending on the material. Here, through rigorous first-principles calculations, we uncover a phonon-dominated thermal transport regime in the Weyl semimetals TaAs and TaP. Remarkably, despite its metallic character, TaP exhibits an exceptionally high phonon thermal conductivity ($\kappa_{\rm ph}$) of 162 Wm$^{-1}$K$^{-1}$ at room temperature, surpassing its electronic counterpart by nearly an order of magnitude. This anomalously high $\kappa_{\rm ph}$ arises from the combined effects of Weyl electronic structure, acoustic phonon bunching, and a large low–high frequency phonon gap, which together suppress phonon-electron and three-phonon scattering. The linearly dispersing bands near the Fermi level yield a low electronic density of states, limiting both electrical conductivity and electronic thermal transport in these compounds. By further surveying a broad range of topological semimetals, we show that the prominence of phonon thermal transport is a universal characteristic of this material class. Our work provides deeper insight into thermal transport mechanisms in topological semimetals and broadens the scope for discovering metals with high thermal conductivity.
\end{abstract}

\maketitle

\section{INTRODUCTION}
Weyl semimetals (WSMs), known for hosting robust Weyl quasiparticles, have garnered significant attention in condensed matter physics and materials research \cite{wan2011topological,yan2017topological, armitage2018weyl, ong2021experimental,PhysRevX.5.011029,lv2015experimental,yang2015weyl}. Distinguished by their linear bulk band dispersion relations and the presence of robust Weyl points, WSMs exhibit a range of intriguing phenomena such as Fermi arc surface states \cite{xu2015observation,sun2015topological}, high mobility \cite{shekhar2015extremely}, large unsaturated magnetic moments \cite{shekhar2015large,du2016large,zhang2019non}, negative magneto-resistance \cite{arnold2016negative, PhysRevX.5.031023,li2017negative}, and chiral anomaly \cite{ashby2014chiral,PhysRevX.5.031023,jia2016weyl,yuan2020discovery}. Among the prototypical WSMs, TaAs family (TaAs, TaP, NbAs, and NbP) has been well studied, with experiments directly confirming the existence of Weyl fermions and Fermi‑arc surface states, thereby establishing its non‑trivial topological nature \cite{PhysRevX.5.011029,xu2016optical, lv2015experimental, PhysRevX.5.031023, yang2015weyl}. While the electronic and topological properties of WSMs have been extensively explored, their thermal transport characteristics remains largely uncharted. Given its importance for both fundamental science and diverse applications, establishing a comprehensive understanding of heat conduction in such systems is imperative. Moreover, in Weyl semimetals, the linearly dispersing bands near the Fermi level typically lead to low carrier densities and consequently weak phonon–electron (ph-el) scattering \cite{PhysRevB.110.054304,kundu2021ultrahigh,jin2025anharmonicity}, which facilitates an enhancement of the phonon contribution to thermal conductivity. These features suggest that WSMs may provide a promising platform for achieving high thermal conductivity.

In many metals, electronic thermal conductivity ($\kappa_{\rm e}$) is often the dominant contribution to heat transport, although the lattice contribution can remain non-negligible depending on the material. In such cases, the electronic thermal conductivity can be estimated from electrical conductivity ($\sigma$) via the Wiedemann–Franz law (WFL), expressed as $\kappa_{\rm e} = L\sigma T$ \cite{chester1961law}. Here, the Lorenz number $L$ is commonly taken as the Sommerfeld value \cite{makinson1938thermal,klemens1986thermal}, $L_{0} = \pi^{2} k_{\rm B}^{2}/3e^{2} = 2.445 \times 10^{-8}$ W$\Omega$K$^{-2}$. However, this picture may not hold in WSMs, where the low carrier density associated with linearly dispersing bands near the Fermi level can substantially suppress $\kappa_{\rm e}$. At the same time, weak ph–el scattering, together with favorable phonon band structures, may render the phonon thermal conductivity ($\kappa_{\rm ph}$) substantial or even dominant. Indeed, unusually large $\kappa_{\rm ph}$ have been predicted in several systems, including W \cite{PhysRevB.99.020305}, Be \cite{PhysRevB.109.L220302}, $\theta$-TaN \cite{kundu2021ultrahigh}, and others \cite{PhysRevB.110.054304,PhysRevB.100.144306}. Most remarkably, recent experiments have confirmed that the topological semimetal $\theta$-TaN exhibits an ultrahigh room-temperature (RT) phonon thermal conductivity of approximately 1100 Wm$^{-1}$K$^{-1}$ \cite{doi:10.1126/science.aeb1142,10.1093/nsr/nwag106}, redefining the upper limits of thermal conductivity in metals. These findings demonstrate that phonons can play a dominant role in heat transport even in metallic systems and highlight topological semimetals as a promising platform for discovering materials with exceptionally high thermal conductivity.

In the TaAs family, the low electronic density of states near the Fermi level implies weak ph-el scattering, thereby amplifying the role of phonon thermal transport. Concurrently, the large cation–anion mass difference and strong covalent bonding give rise to a pronounced phonon gap between acoustic and high-frequency optical branches, particularly in TaP \cite{ding2025concurrent}. Although several low-frequency optical branches overlap and intertwine with the acoustic branches, this bunched low-frequency manifold restricts three-phonon (3ph) scattering channels. Collectively, these features substantially reduce the phase space for phonon-phonon scattering processes, particularly acoustic–acoustic–optical scattering ($aao$) channels, thus favoring high phonon thermal conductivity. These characteristics make the TaAs family a promising platform for exploring high phonon thermal conductivity in metallic systems.

In this work, we present a systematic first-principles study of phonon and electron thermal conductivity in the Weyl semimetals TaAs and TaP. Our results reveal that heat conduction in both WSMs is dominated by phonons. In particular, TaP exhibits an exceptionally high phonon thermal conductivity, reaching 162 Wm$^{-1}$K$^{-1}$ along the $a$ axis and 74 Wm$^{-1}$K$^{-1}$ along the $c$ axis at RT, far exceeding its electronic counterpart of 19 and 6 Wm$^{-1}$K$^{-1}$, respectively. This phonon-dominated thermal transport behavior arises from the combined effects of weak ph-el scattering and suppressed phonon–phonon scattering. The former originates from the low electronic density of states near the Fermi level associated with the Weyl electronic structure, whereas the latter is facilitated by acoustic phonon bunching and a large low–high frequency phonon gap. Furthermore, the increased mass difference from TaAs to TaP enlarges the phonon gap and contributes to the pronounced enhancement of phonon thermal conductivity in TaP. These findings highlight the crucial role of phonons in heat transport in topological semimetals.

\section{METHODOLOGY}
\subsection{Phonon thermal transport}
In metals, the total $\kappa$ comprises contributions from both phonons and electrons, i.e., $\kappa=\kappa_{\rm ph}+\kappa_{\rm e}$. The phonon thermal conductivity tensor $\kappa_{\rm ph}^{\alpha \beta}$ is obtained by solving the linearized phonon Boltzmann transport equation (BTE), which can be expressed as
\begin{equation}
    \kappa_{\rm ph}^{\alpha \beta} = \sum_{p \mathbf{q}} C_{p \mathbf{q}}{\rm v}^{\alpha}_{p \mathbf{q}} \otimes {\rm F}^{\beta}_{p \mathbf{q}},
\end{equation}
where $p$, $\mathbf{q}$, $C_{p\mathbf{q}}$, v$^{\alpha}_{p \mathbf{q}}$, and F$^{\beta}_{p \mathbf{q}}$ are the phonon branches, wave vector, mode-resolved heat capacity, phonon group velocity, and mean free path, respectively, with $\alpha$ and $\beta$ representing Cartesian axes. F$_{p \mathbf{q}}$ is limited by multiple scattering mechanisms, including 3ph (1/$\tau^{\rm 3ph}_{p \mathbf{q}}$), four-phonon (4ph, 1/$\tau^{\rm 4ph}_{p \mathbf{q}}$), isotope (ph-iso, 1/$\tau^{\rm iso}_{p \mathbf{q}}$), and ph-el (1/$\tau^{\rm ph-el}_{p \mathbf{q}}$) scattering processes. Detailed expressions for 3ph, 4ph, and ph-iso can be found elsewhere \cite{wei2024tensile,PhysRevB.110.054304}. The ph-el scattering is obtained from the imaginary part of the phonon self-energy $\Pi_{p \mathbf{q}}^{\prime\prime}$ \cite{PhysRevLett.121.175901}, given by $1/\tau_{p \mathbf{q}}^{\rm ph-el} = 2\Pi_{p \mathbf{q}}^{\prime\prime}/\hbar$, which can be directly obtained from the phonon linewidth computed using the EPW code \cite{ponce2016epw}.

\subsection{Phonon-limited electrical transport}
Under an external electric field, the electrical conductivity ($\sigma^{\alpha \beta}$) tensor can be obtained by solving the electron BTE, given as\cite{zhou2021perturbo}
\begin{equation}\label{eleSigma}
\sigma^{\alpha \beta} = \frac{2e^{2}}{N_{\mathbf{k}}V_0 k_{\rm B} T} \sum_{n\mathbf{k}} (-\frac{\partial f_{nk}^{0}}{\partial \epsilon_{nk}}) \mathbf{\nu}^{\alpha}_{n\mathbf{k}} \otimes \mathbf{F}^{\beta}_{n\mathbf{k}},
\end{equation} 
where $N_{\mathbf{k}}$, $e$, $V_{0}$, $k_{\rm B}$, $n$, $\mathbf{k}$, and $f^{0}_{n\mathbf{k}}$ denote the number of uniformly sampled $\mathbf{k}$ points, volume, elementary charge, Boltzmann constant, band index, wave vector and the equilibrium Fermi–Dirac distribution, respectively. Within the relaxation time approximation (RTA), $\mathbf{F_{nk}^{\beta}}$ = $\tau_{nk}v_{nk}^{\beta}$ represents the electronic mean free path. To obtain more accurate results, we solve Eq.~\ref{eleSigma} iteratively \cite{li2015electrical,PhysRevB.94.085204} using the RTA results as the initial guess. When the thermal gradient is further included, Eq.~\ref{eleSigma} can be written as
\begin{equation}
\sigma^{\alpha\beta} = e^2 \int dE \left(-\frac{\partial f^0}{\partial E}\right) \Sigma^{\alpha\beta}(E).
\end{equation}
Here, $\Sigma^{\alpha\beta}(E)$ is the transport distribution function, which is computed using the tetrahedron integration method \cite{PhysRevB.49.16223}, written as 
\begin{equation}
    \Sigma^{\alpha\beta}(E) = \frac{2}{N_k V k_{B} T} \sum_{nk} v_{nk}^\alpha F_{nk}^\beta \delta(E - \varepsilon_{nk}).
\end{equation}
Then, the Seebeck coefficient ($S^{\alpha \beta}$) and electrical thermal conductivity ($\kappa_{\rm e}$) can be obtained as 
\begin{equation}
S^{\alpha \beta} = \frac{e}{T}
\frac{
\displaystyle \int dE \, (E - \mu) \left( - \frac{\partial f^0}{\partial E} \right) \Sigma^{\alpha \beta}(E)
}{\sigma^{\alpha \beta}},
\end{equation}

\begin{equation}\label{kappae}
\begin{split}
\kappa_{\rm e}^{\alpha\beta} = K^{\alpha\beta} - T (\sigma^{\alpha\beta}S^{\alpha\beta})^2/\sigma^{\alpha\beta},
\end{split}
\end{equation}
where $\mu$ is the chemical potential, and $K^{\alpha \beta}$ denotes the electronic contribution to the thermal conductivity under zero electric-field conditions, which can be expressed as
\begin{equation}
K^{\alpha\beta} = \frac{1}{T} \int dE \, (E - \mu)^2 \left( -\frac{\partial f^0}{\partial E} \right) \Sigma^{\alpha\beta}(E).
\end{equation}

\begin{figure*}[t]
	\includegraphics[width=\linewidth]{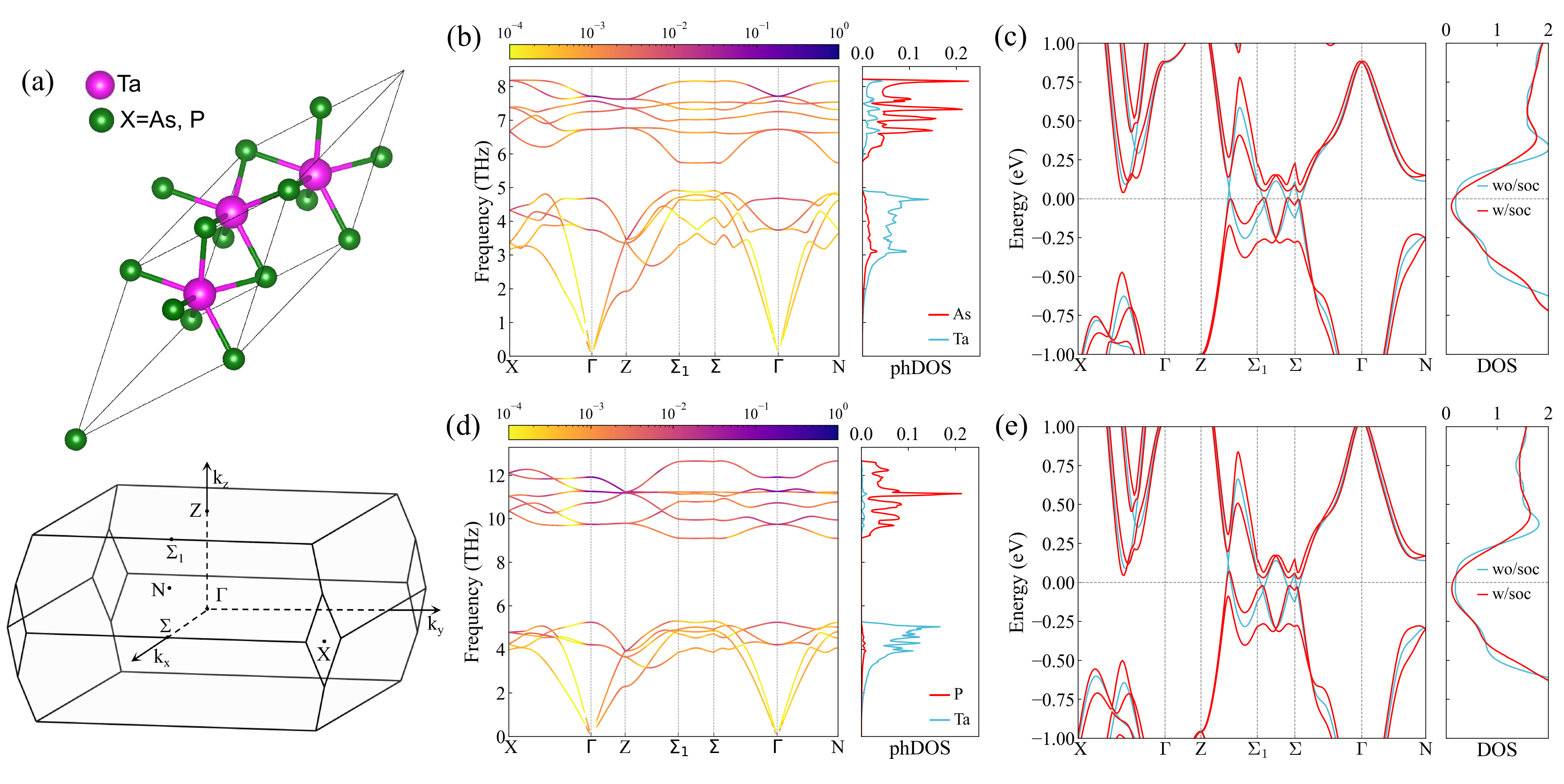}
	\caption{(a) The primitive cell and Brillouin Zone of Ta$X$ ($X$ = As, P). Phonon dispersion relation and projected phonon density of states (phDOS) of (b) TaAs and (d) TaP, where the phonon linewidth due to ph-el scattering is projected onto the corresponding phonon structure. Electronic band structure and electronic DOS of (c) TaAs and (e) TaP without and with including SOC.}
\label{fig1}
\end{figure*}

\subsection{Computational details}
The phonon dispersions and anharmonic interatomic force constants (IFCs) were calculated within density functional theory (DFT) using the Vienna \textit{Ab Initio} Simulation Package (VASP) \cite{kresse1993ab,kresse1996efficient} with the projector augmented-wave (PAW) method \cite{blochl1994projector}. Exchange–correlation effects were treated within the generalized gradient approximation (GGA) using the Perdew–Burke–Ernzerhof (PBE) functional \cite{perdew1996generalized}. The second-, third-, and fourth-order IFCs were obtained using the finite-displacement supercell approach with a $3\times3\times3$ supercell. Third- and fourth-order IFCs were truncated at cutoff radii of 0.5 and 0.3 nm, respectively. Electron–phonon (el–ph) coupling calculations were performed within density functional perturbation theory (DFPT) using the Quantum ESPRESSO package \cite{giannozzi2009quantum}. Fully relativistic ultrasoft pseudopotentials were employed to account for spin–orbit coupling (SOC). The kinetic-energy cutoffs for the wave functions and charge density were set to 80 and 800 Ry, respectively. Self-consistent calculations were carried out on a $12\times12\times12$ $\mathbf{k}$-point mesh with a $4\times4\times4$ $\mathbf{q}$-point grid, using a total-energy convergence threshold of $10^{-12}$ eV. The el–ph matrix elements were first computed on the coarse $12\times12\times12$ $\mathbf{k}$-point grids and subsequently interpolated onto dense $60\times60\times60$ $\mathbf{k}$-point meshes using maximally localized Wannier functions as implemented in the EPW package \cite{ponce2016epw}. The resulting phonon linewidths and ph–el scattering rates were incorporated into the phonon BTE. The $\kappa_{\rm ph}$ was then obtained by iteratively solving the phonon BTE using a modified version of ShengBTE \cite{li2014shengbte,han2022fourphonon,yang2021indirect,yang2021tuning}, which explicitly includes ph–el scattering. To evaluate phonon-limited electronic transport properties, the el–ph scattering rates were further interpolated onto dense $80\times80\times80$ $\mathbf{k}$- and $\mathbf{q}$-point grids and the electron BTE was solved using the Perturbo package \cite{zhou2021perturbo}. Additional convergence tests and computational details are provided in the Supplemental Material (SM) \cite{SM}.

\section{RESULTS AND DISCUSSION}
\subsection{Electronic and phonon band structures}
TaAs and TaP both crystallize in a body-centered tetragonal structure with space group $I4_{1}md$ (No. 109), with each primitive cell containing four atoms, as illustrated in Fig.~\ref{fig1}(a). The optimized lattice parameters of TaAs (TaP) are $a=b=3.456$ (3.329) $\rm \AA$ and $c=11.718$ (11.392) $\rm \AA$, in good agreement with previous experimental and theoretical reports \cite{PhysRevX.5.031023,PhysRevX.5.011029,lv2015observation,lv2015experimental,x8zl-w5x3}. Both compounds exhibit pronounced structural anisotropy, which is reflected in a variety of physical properties, including their electronic and thermal transport behavior.

Figures~\ref{fig1}(b) and (d) display the phonon dispersions along high-symmetry directions in the irreducible Brillouin zone together with the projected phonon density of states (phDOS). Replacing As with P shifts the entire phonon spectrum toward higher frequencies, consistent with the stronger bonding strength in TaP reported previously \cite{ding2025concurrent}. 
Both compounds exhibit a pronounced phonon gap separating the acoustic branches from the high-frequency optical branches. This gap is more substantial in TaP, owing to its larger cation–anion mass difference relative to TaAs. Such a gap restricts available $aao$ scattering channels due to the constraints of energy selection rules. We also note that the low-frequency acoustic branches in TaP are closely bunched together, which further limits the phase space available for three-acoustic-phonon ($aaa$) scattering processes under the constraints of energy selection rules \cite{lindsay2013first}. These features  suppress 3ph scattering involving heat-carrying acoustic modes and consequently contribute to the high phonon thermal conductivity of TaP. 

The electronic band structures of TaAs and TaP with and without SOC are presented in Figs.~\ref{fig1}(c) and (e). In the absence of SOC, the valence and conduction bands intersect near the Fermi level, forming mirror-symmetry-protected nodal rings within the mirror plane \cite{PhysRevX.5.011029}. When SOC is included, these nodal rings are gapped out along high-symmetry directions, and the band degeneracies are lifted except at isolated crossing points, resulting in the formation of Weyl nodes. This evolution of band topology has been extensively discussed in previous studies \cite{PhysRevX.5.011029,lee2015fermi}. Since the electronic structure near the Fermi level plays a crucial role in determining transport properties, SOC is explicitly included in all subsequent calculations.

A notable consequence of the Weyl electronic structure is the extremely low electronic density of states (DOS) at the Fermi level, as shown in the right panels of Figs.~\ref{fig1}(c) and (e). The reduced DOS leads to a low carrier concentration and intrinsically weak el–ph coupling, since the scattering strength generally scales with the electronic DOS at the Fermi level \cite{kundu2021ultrahigh}. This picture is further supported by the calculated ph–el scattering rates projected onto the phonon dispersions [Figs.~\ref{fig1}(b) and (d)], which remain remarkably small throughout the spectrum. In particular, the scattering rates of acoustic phonons are below 0.01 ps$^{-1}$, indicating that ph–el scattering processes provide only a minor resistance to heat transport in these two Weyl semimetals.

\begin{figure}[btp]
	\includegraphics[width=1.02\linewidth]{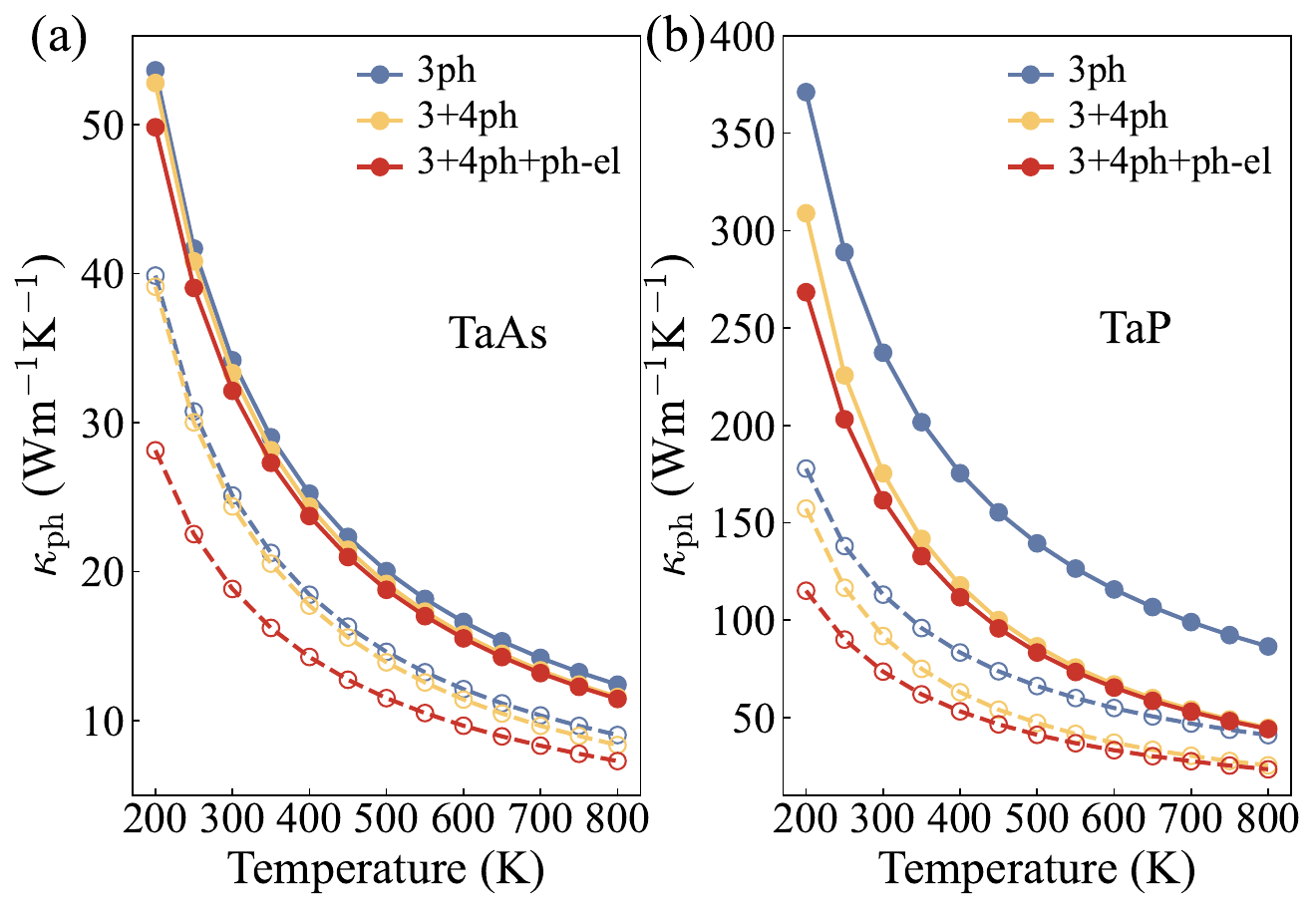}
	\caption{Temperature-dependent $\kappa_{\rm{ph}}$ of (a) TaAs and (b) TaP with different combinations of scattering mechanisms. The filled and empty symbols correspond to the $\kappa_{\rm ph}$ along the $a$ and $c$ axes, respectively.}
\label{fig2}
\end{figure}

\subsection{Phonon thermal conductivity}
The intrinsic lattice thermal conductivity of TaAs and TaP is calculated over the temperature range of 200–800 K by incorporating 3ph, 4ph, and ph–el scattering processes. The results are summarized in Fig.~\ref{fig2} and compared with available theoretical and experimental data.

For TaAs, considering only 3ph scattering, Peng \textit{et al}. predicted a RT $\kappa_{\rm ph}$ of 36 Wm$^{-1}$K$^{-1}$ \cite{PENG2016225}, while Ouyang \textit{et al}. reported a similar value of 39 Wm$^{-1}$K$^{-1}$ \cite{C6CP02935C}. Both values are substantially higher than the experimental value of $\sim$19 Wm$^{-1}$K$^{-1}$ \cite{xiang2017anisotropic}, which was extracted indirectly from the measured thermal conductivity by subtracting the electronic contribution estimated using the WFL. As discussed by Han \textit{et al}. \cite{HAN2023520}, the discrepancy between theory and experiment may arise from carrier-dependent ph–el scattering, sample quality, defects, and uncertainties associated with the assumed Lorenz number. Our calculated RT value of 34 Wm$^{-1}$K$^{-1}$, obtained by considering only 3ph scattering, agrees well with previous first-principles studies, supporting the reliability of the present computational approach.

We then examine the role of higher-order anharmonicity. The influence of 4ph scattering differs markedly between the two Weyl semimetals. For TaAs, including 4ph scattering results in only a modest reduction of $\kappa_{\rm ph}$ throughout the entire temperature range. In contrast, TaP exhibits a much stronger 4ph effect, with $\kappa_{\rm ph}$ reduced by approximately 26\% at RT. This pronounced difference can be traced to the evolution of the phonon spectrum from TaAs to TaP. The larger mass difference between Ta and P increases the phonon gap between the low- and high-frequency branches. Together with the bunched low-frequency acoustic modes, the wider gap suppresses 3ph scattering channels and thereby enhances the relative importance of 4ph processes in TaP.

Besides, ph–el scattering also affects heat transport, and its impact is highly anisotropic. Along the $a$ axis, the inclusion of ph–el scattering brings in only a negligible change in $\kappa_{\rm ph}$ for both compounds. Along the $c$ axis, however, the reduction is substantial. For TaAs, $\kappa_{\rm ph}$ decreases by approximately 23\% at RT upon including ph–el scattering. This anisotropic behavior is consistent with the mode-resolved ph–el scattering rates shown in Figs.~\ref{fig1}(b) and (d), where acoustic phonons propagating along the $\Gamma$–Z direction exhibit significantly stronger ph–el scattering than those along other high-symmetry directions.

When all scattering mechanisms are taken into account, TaP exhibits an exceptionally high RT phonon thermal conductivity of 162 Wm$^{-1}$K$^{-1}$ along the $a$ axis, far exceeding the corresponding value of 32 Wm$^{-1}$K$^{-1}$ in TaAs. The higher $\kappa_{\rm ph }$ in TaP mainly originates from its elevated phonon frequencies and wider phonon gap between the low- and high-frequency branches, which substantially reduce the available 3ph scattering phase space and lead to weaker 3ph scattering compared with TaAs. Furthermore, both compounds exhibit pronounced thermal-conductivity anisotropy, with the in-plane $\kappa_{\rm ph}$ substantially exceeding that along the $c$ axis, arising from their strongly anisotropic crystal structures.

\begin{figure}[!btp]
	\includegraphics[width=\linewidth]{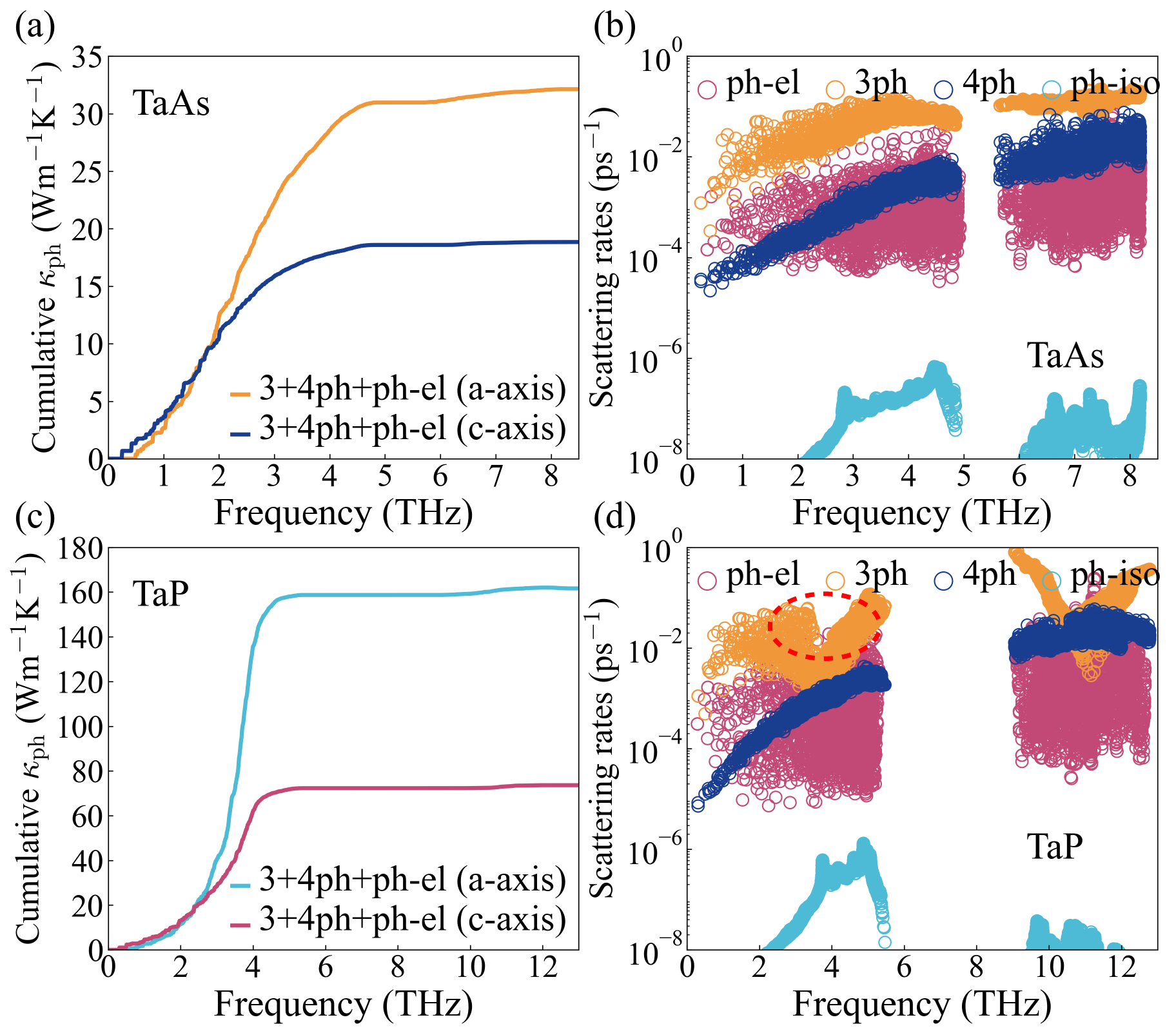}
	\caption{Calculated cumulative $\kappa_{\rm ph}$ of (a) TaAs and (c) TaP along the $a$-axis at RT. The phonon scattering rates contributed from 3ph, 4ph, ph-iso, and ph-el processes for (b) TaAs and (d) TaP at RT.}
\label{fig3}
\end{figure}

To clarify the frequency ranges affected by different scattering mechanisms, we plot in Figs.~\ref{fig3}(a) and (c) the cumulative contributions to $\kappa_{\rm ph}$ along the $a$-axis for TaAs and TaP at RT. In both compounds, heat transport is dominated by phonons below 5 THz, which correspond primarily to acoustic modes. For TaAs, the contribution of these heat-carrying phonons is only weakly affected by either 4ph or ph–el scattering. As shown in Fig.~\ref{fig3}(b), the 3ph scattering rates in this frequency range exceed the corresponding 4ph and ph–el scattering rates by nearly two orders of magnitude, indicating that heat resistance remains predominantly governed by 3ph processes.

The situation is markedly different in TaP. As shown in Fig.~\ref{fig3}(c), the contribution of phonons in the 3–5 THz range is substantially suppressed upon inclusion of 4ph scattering. This behavior can be traced to the pronounced dip in the 3ph scattering rates near 4 THz, as highlighted by the oval in Fig.~\ref{fig3}(d). This reduction is enabled by the distinctive phonon spectrum of TaP, which combines a large separation between the low- and high-frequency branches with pronounced bunching of acoustic branches [Fig.~\ref{fig1}(d)]. These features significantly reduce the phase space available for 3ph processes, as seen in Fig. S2 \cite{SM}, particularly $aao$ and $aaa$ scattering channels \cite{chen2025symmetry,wang2025atomic}. As a consequence, the 3ph scattering rates are strongly suppressed, rendering the 4ph scattering rates comparable in magnitude within this frequency window. This explains the much stronger influence of 4ph scattering on $\kappa_{\rm ph}$ in TaP than in TaAs.

By contrast, the ph–el scattering rates remain substantially smaller than the 3ph scattering rates throughout most of the spectrum in both compounds. Although ph–el scattering contributes to the reduction of $\kappa_{\rm ph}$, particularly along the $c$ axis, its overall effect is secondary compared with intrinsic phonon-phonon scattering. In addition, ph–iso scattering is negligibly weak in both TaAs and TaP and has virtually no impact on the resulting thermal conductivity.

\subsection{Phonon-limited electrical transport}
\begin{figure*}[!btp]
	\includegraphics[width=\linewidth]{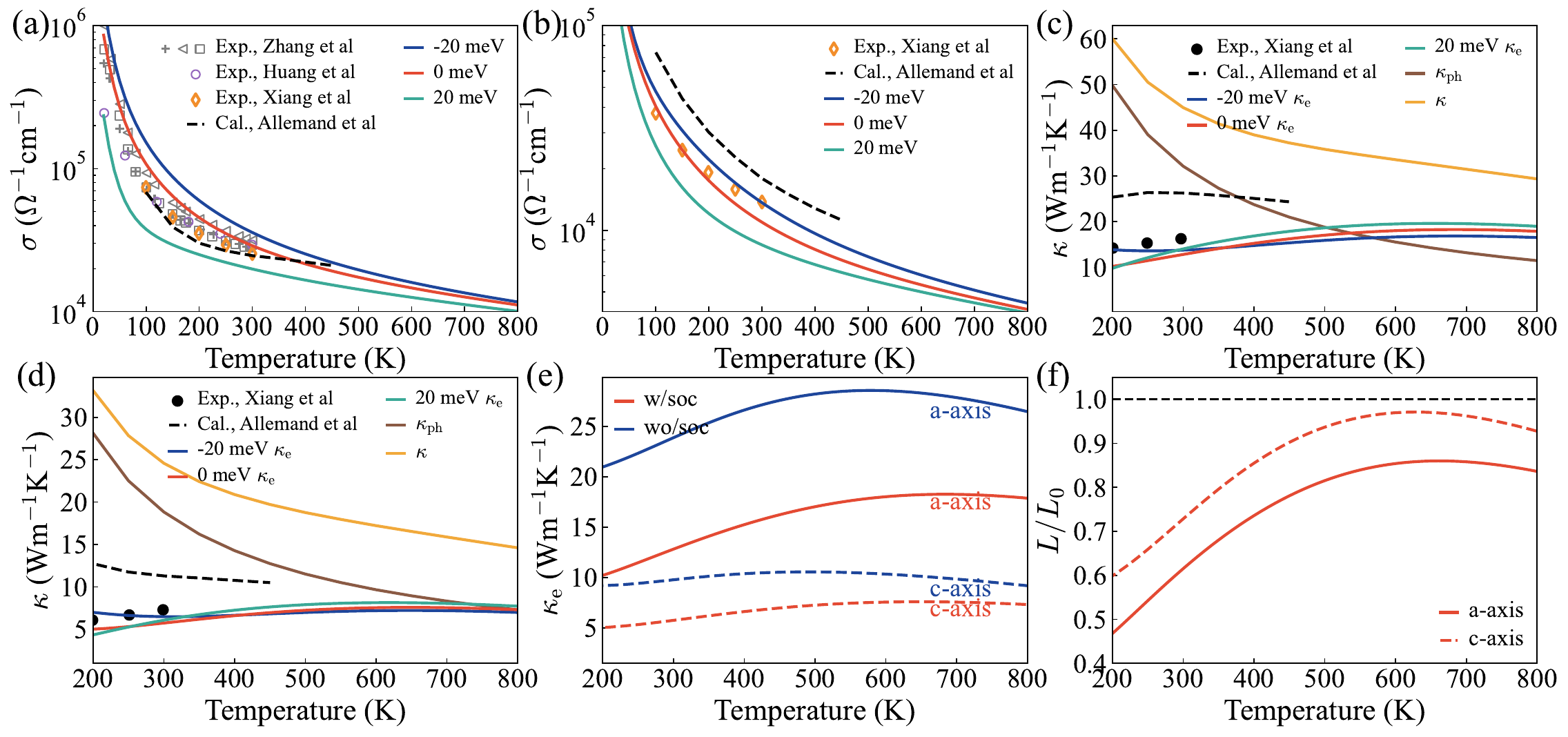}
	\caption{The electrical conductivity $\sigma$ of TaAs along the (a) $a$-axis and (b) $c$-axis. The phonon thermal conductivity $\kappa_{\rm ph}$, electronic thermal conductivity $\kappa_{\rm e}$, and total thermal conductivity $\kappa = \kappa_{\rm ph} + \kappa_{\rm e}$ of TaAs along the (c) a-axis and (d) c-axis. The hollow symbols represent the experimental data reported in Refs. \cite{witczak2012topological, zhang2017electron, xiang2017anisotropic}, while the black dotted line denotes the theoretical results calculated by Allemand \textit{et al}. \cite{7lnq-snmp}. In addition, the electrical transport properties within an energy range of ±20 meV around the Fermi level are also presented. (e) $\kappa_{\rm e}$ of TaAs with and without SOC along the $a$- and $c$-axis. (f) Temperature dependence of the Lorenz number in TaAs, compared with the theoretical data obtained from Allemand \textit{et al}.~\cite{7lnq-snmp}.}
\label{fig4}
\end{figure*}

\begin{figure*}[!btp]
	\includegraphics[width=\linewidth]{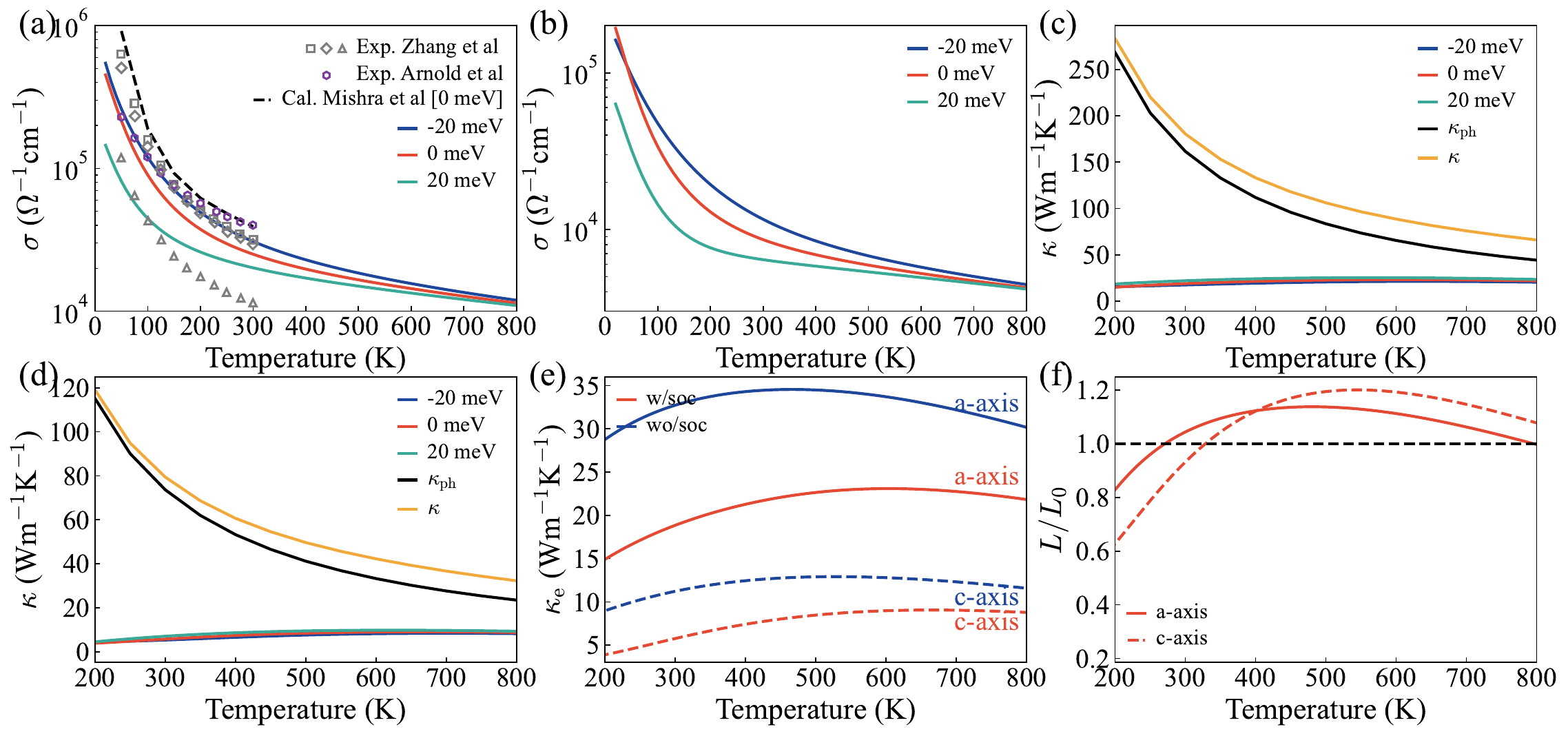}
	\caption{The electrical conductivity $\sigma$ of TaP along the (a) a-axis and (b) c-axis. The phonon thermal conductivity $\kappa_{\rm ph}$, electronic thermal conductivity $\kappa_{\rm e}$, and total thermal conductivity $\kappa = \kappa_{\rm ph} + \kappa_{\rm e}$ of TaP along the (c) a-axis and (d) c-axis. In addition, the electrical transport properties within an energy range of ±20 meV around the Fermi level are also presented. The hollow symbols represent the experimental data reported in Refs. \cite{arnold2016negative,PhysRevB.92.041203}, while the black dotted line represents the theoretical data calculated by Mishra et al \cite{d9np-11j1} with no Fermi level shift (0 meV). (e) $\kappa_{\rm e}$ of TaP calculated with and without SOC. (f) Temperature dependence of the Lorenz number in TaP.
}
\label{fig5}
\end{figure*}

Having established the phonon thermal transport properties of TaAs and TaP, we now turn to their electronic transport behavior. The calculated phonon-limited electrical conductivities and electronic thermal conductivities are presented in Figs.~\ref{fig4} and \ref{fig5}. Since the carrier concentration in real samples can be readily altered by impurities, defects, and unintentional doping, the Fermi level may deviate from its intrinsic position. Given the strong sensitivity of transport properties to the Fermi level in WSMs, we further evaluate the electrical transport coefficients within an energy window of $\pm$20 meV around the Fermi level to facilitate a more meaningful comparison with experimental measurements.

Figure~\ref{fig4}(a) shows the temperature-dependent electrical conductivity of TaAs along the $a$ axis. At the intrinsic Fermi level, our calculated conductivity agrees reasonably with both previous experimental measurements and theoretical predictions \cite{witczak2012topological,zhang2017electron,xiang2017anisotropic,7lnq-snmp}. As expected, shifting the Fermi level by $\pm$20 meV mainly results in an overall increase or decrease in the conductivity while preserving its temperature dependence. A similar comparison is presented for the $c$ axis in Fig.~\ref{fig4}(b). In this direction, our calculated intrinsic conductivity is in good agreement with the experimental data of Xiang \textit{et al}. \cite{xiang2017anisotropic}, and the previous theoretical results \cite{7lnq-snmp} somewhat overestimate the conductivity across the entire temperature range.

Figures~\ref{fig4}(c) and (d) compare our predicted $\kappa_{\rm ph}$, $\kappa_{\rm e}$, and $\kappa_{\rm tot}$ with available literature data along the $a$ and $c$ axes, respectively. Xiang \textit{et al}. $\cite{xiang2017anisotropic}$ estimated $\kappa_{\rm e}$ from the measured electrical resistivity using the WFL, assuming a constant Lorenz number equal to the Sommerfeld value. Their estimated $\kappa_{\rm e}$ is slightly higher than our calculated values. In comparison, the first-principles calculations by Allemand \textit{et al}. \cite{7lnq-snmp} yield $\kappa_{\rm e}$ values along both axes that are significantly larger than our results and substantially exceed the available experimental estimates. The discrepancy between the two predictions may primarily originate from differences in the computational methodologies, particularly the choice of pseudopotentials. In the work of Allemand \textit{et al}., norm-conserving pseudopotentials were employed, whereas ultrasoft pseudopotentials were used in our calculations. Such differences can lead to variations in the electronic structure, especially near the Fermi level, thereby affecting the calculated transport properties. This effect can be particularly important for Weyl semimetals, where electronic transport is highly sensitive to the detailed band structure and carrier characteristics in the vicinity of the Fermi energy. More importantly, our results show that phonons dominate the thermal transport of TaAs in the low- and intermediate-temperature regimes. Along the $a$ axis, $\kappa_{\rm ph}$ reaches approximately 32 Wm$^{-1}$K$^{-1}$ at RT, whereas $\kappa_{\rm e}$ is only about 13 Wm$^{-1}$K$^{-1}$. A similar behavior is observed along the $c$ axis.

To more accurately predict the electronic thermal transport, Fig.~\ref{fig4}(e) compares the calculated $\kappa_{\rm e}$ with and without SOC. SOC is found to play a crucial role in determining the electronic thermal conductivity, reducing $\kappa_{\rm e}$ substantially along both axes. In contrast, the influence of SOC on the electrical conductivity is relatively small, as demonstrated in the Supplemental Material \cite{SM}. This distinct behavior reflects the strong sensitivity of heat-carrying electronic states to the SOC-induced modifications of the band structure near the Fermi level.

Based on the calculated intrinsic $\sigma$ and $\kappa_{\rm e}$, we further evaluate the Lorenz number, $L=\kappa_{\rm e}/(\sigma T)$, as shown in Fig.~\ref{fig4}(f). Remarkably, the Lorenz number remains substantially below the Sommerfeld value, $L_0$, throughout the entire temperature range, with the deviation being particularly pronounced at lower temperatures. Moreover, the reduction is stronger along the $a$ axis than along the $c$ axis, revealing significant anisotropy in the el-ph scattering processes. In contrast, Allemand \textit{et al}. \cite{7lnq-snmp} predicted a Lorenz number with a substantial upward deviation from the Sommerfeld value, reaching $\sim$70\% above the Sommerfeld limit. This behavior is qualitatively different from our results and mainly originates from their significantly larger calculated $\kappa_{\rm e}$, as discussed above. According to previous theoretical analyses \cite{PhysRevB.102.174306}, such suppressed Lorenz numbers arise from inelastic el–ph scattering. Unlike electrical transport, which is mainly relaxed by large-angle scattering events, thermal transport can be relaxed by both large-angle and small-angle scattering processes. Consequently, the energy relaxation time becomes substantially shorter than the momentum relaxation time, leading to a Lorenz number well below $L_0$. The pronounced violation of the WFL indicates a substantial decoupling of charge and heat transport in TaAs. Combined with its high carrier mobility \cite{PhysRevX.5.031023}, this feature suggests that TaAs may represent a promising platform for low-temperature thermoelectric applications.

Figure~\ref{fig5} presents the calculated electrical and thermal transport properties of TaP together with the recently reported first-principles and experimental results \cite{d9np-11j1,arnold2016negative,PhysRevB.92.041203}. Similar to TaAs, the electrical conductivity exhibits a noticeable sensitivity to the Fermi level. Our calculated $\sigma$ is close to the previous prediction, and both calculations reasonably reproduce the experimental measurements \cite{d9np-11j1}, given the strong dependence of carrier concentration on sample quality. As in TaAs, the discrepancy between the two theoretical calculations may primarily originate from the different choices of pseudopotentials, which can lead to variations in the electronic structure near the Fermi level and consequently affect the calculated transport properties.

\begin{figure*}[!btp]
	\includegraphics[width=0.9\linewidth]{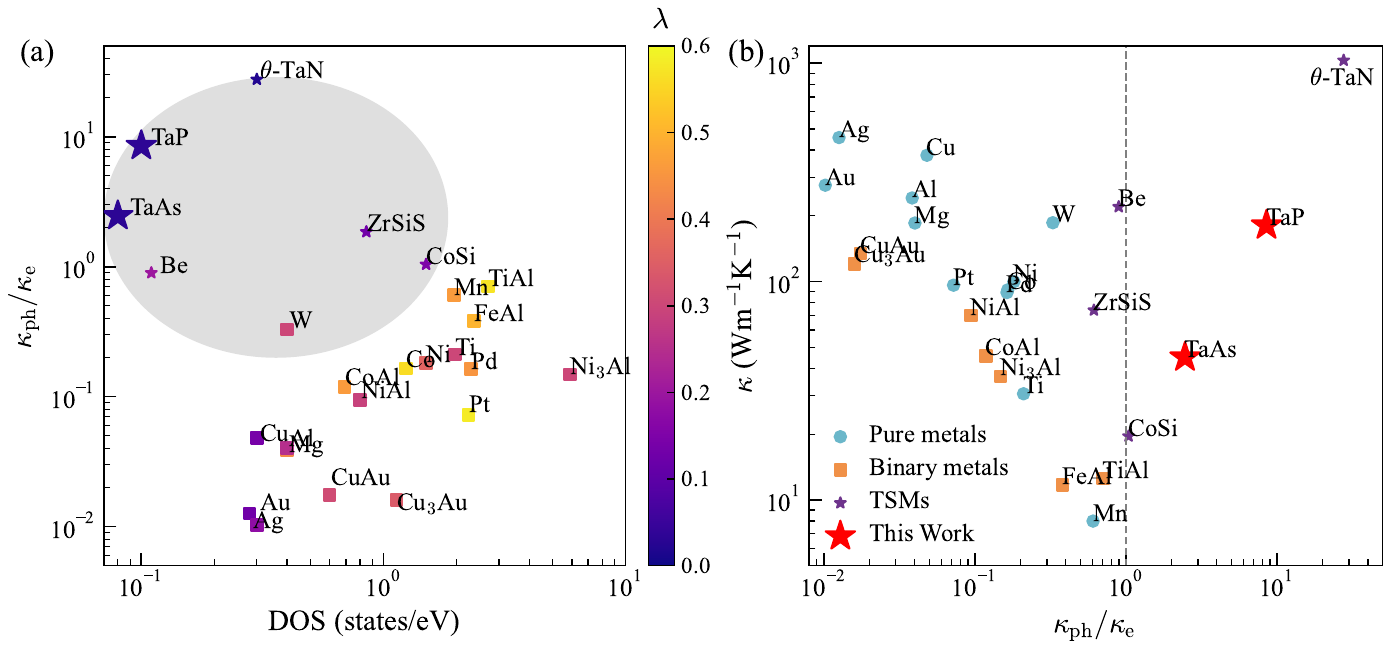}
	\caption{(a) The relative importance of $\kappa_{\rm ph}$, as gauged by the ratio $\kappa_{\rm ph}$/$\kappa_{\rm e}$, for available topological semimetals and metals; the colorbar represents the corresponding el-ph coupling strength ($\lambda$). (b) The total $\kappa$ versus the ratio of $\kappa_{\rm ph}$/$\kappa_{\rm e}$ for available topological semimetals and metals. The presented data for the electronic DOS, $\kappa_{\rm ph}$, $\kappa_{\rm e}$, and $\lambda$ are extracted from Refs. \cite{PhysRevB.100.144306,kundu2021ultrahigh,PhysRevLett.123.136802} and the Material Project database \cite{10.1063/1.4812323,munro2020improved}. Topological semimetals are marked with stars.}
\label{fig6}
\end{figure*}

The corresponding $\kappa_{\rm ph}$, $\kappa_{\rm e}$, and total thermal conductivity are displayed in Figs.~\ref{fig5}(c) and (d). Compared with TaAs, the dominance of phonon heat transport is considerably more pronounced in TaP. Throughout the entire temperature range investigated, $\kappa_{\rm ph}$ remains substantially larger than $\kappa_{\rm e}$ along both axes. At RT, $\kappa_{\rm ph}$ reaches approximately 162 Wm$^{-1}$K$^{-1}$ along the $a$ axis and 74 Wm$^{-1}$K$^{-1}$ along the $c$ axis, whereas the corresponding electronic contributions are only about 19 and 6 Wm$^{-1}$K$^{-1}$, respectively. Consequently, the lattice thermal conductivity exceeds the electronic contribution by nearly an order of magnitude, establishing TaP as a prototypical phonon-dominated metal. Figure~\ref{fig5}(e) also compares the calculated $\kappa_{\rm e}$ with and without SOC. Similar to TaAs, SOC substantially suppresses $\kappa_{\rm e}$ along both axes, highlighting the role of SOC-induced modifications of the band structure near the Fermi level.

A particularly intriguing feature of TaP is revealed by the temperature dependence of the Lorenz number shown in Fig.~\ref{fig5}(f). Unlike TaAs, TaP exhibits a crossover from $L<L_0$ at low temperatures to $L>L_0$ above RT. The Lorenz number reaches a maximum around 500–600 K, exceeding the Sommerfeld limit by nearly 20\%. Similar behavior has recently been reported in the topological semimetal CoSi \cite{5svq-g1k7}. This anomaly has been attributed to electron–hole compensation driven by the topologically dictated band structure, which gives rise to pronounced bipolar diffusive transport. Together with the exceptionally high $\kappa_{\rm ph}$, this unusual $L$ further underscores the unconventional thermal transport behavior in WSMs.

\subsection{Further insights into phonon-dominated thermal transport}
The above results establish TaAs and TaP as representative examples of phonon-dominated thermal transport in WSMs. Such behavior should not be unique to these two compounds, but appears to be a general characteristic of a broader class of topological semimetals. The underlying origin can be traced to their unique electronic band structures. Owing to the presence of linearly dispersing bands near the Fermi level, topological semimetals typically possess a low electronic DOS at the Fermi level ($N_F$), which limits the available ph-el phase space and consequently leads to weak ph–el scattering. As a result, phonons can contribute substantially, and in some cases dominantly, to heat transport.

To further illustrate this trend, Fig.~\ref{fig6}(a) summarizes the ratio $\kappa_{\rm ph}/\kappa_{\rm e}$ as a function of the electronic DOS at the Fermi level ($N_F$) for reported topological semimetals \cite{PhysRevX.5.011029,jin2025anharmonicity,PhysRevLett.123.136802,PhysRevB.93.241202,rao2019observation}, together with representative elemental and binary metals \cite{PhysRevB.99.020305,PhysRevB.100.144306}. The color scale represents the el–ph coupling strength ($\lambda$). A clear inverse correlation between $\kappa_{\rm ph}/\kappa_{\rm e}$ and $N_F$ is observed. Owing to their nontrivial electronic structures, topological semimetals generally possess a low $N_F$, which is typically associated with weak el–ph coupling. Consequently, they exhibit substantially larger $\kappa_{\rm ph}/\kappa_{\rm e}$ ratios than conventional metals. Among all materials considered here, TaP exhibits one of the largest $\kappa_{\rm ph}/\kappa_{\rm e}$ ratios, surpassed only by $\theta$-TaN, highlighting the exceptionally large phonon contribution to heat transport in this compound.

This trend is further illustrated in Fig.~\ref{fig6}(b), where the total thermal conductivity is plotted as a function of $\kappa_{\rm ph}/\kappa_{\rm e}$. Most topological semimetals are located on the right-hand side of the diagram, corresponding to $\kappa_{\rm ph}/\kappa_{\rm e}>1$, indicating that phonon-dominated heat transport is a widespread characteristic of this material class rather than an isolated phenomenon. In particular, TaAs and TaP occupy the upper-right region of the plot, combining large phonon contributions with high overall thermal conductivity. TaP is especially noteworthy, exhibiting a RT thermal conductivity exceeding 180 Wm$^{-1}$K$^{-1}$. Among the binary metals considered here, this value is second only to that of $\theta$-TaN. Such an exceptionally high thermal conductivity, together with its metallic character, makes TaP a promising candidate for thermal-management applications. More broadly, these results suggest that topological semimetals provide a unique materials platform in which weak el–ph coupling, phonon-dominated heat transport, and ultrahigh thermal conductivity can coexist.

\subsection{Conclusions}
In summary, we have systematically investigated the electrical and thermal transport properties of the Weyl semimetals TaAs and TaP using first-principles calculations combined with Boltzmann transport theory. Our results reveal an unusual phonon-dominated thermal transport regime in both compounds. In particular, TaP exhibits an exceptionally high phonon thermal conductivity of 162 Wm$^{-1}$K$^{-1}$ along the $a$ axis at room temperature, exceeding its electronic counterpart by nearly an order of magnitude. The unusually large phonon contribution originates from the combined effects of weak phonon–electron and phonon–phonon scattering. The former results from the low electronic density of states near the Fermi level, a direct consequence of the topologically nontrivial electronic structure. The latter arises from the cooperative effects of acoustic phonon bunching and the wide phonon gap between acoustic and high-frequency optical branches, which strongly suppress three-phonon scattering processes due to the constraints imposed by energy selection rules. By extending our analysis to a broader range of topological semimetals, we further show that substantial, and often dominant, phonon-mediated heat transport is a common characteristic of this material family. These findings provide new insights into thermal transport in topological semimetals and highlight an effective strategy for identifying metals with exceptionally high thermal conductivity, thereby opening new opportunities for thermal management applications.

\section{acknowledgement}
 This work was supported by the National Natural Science Foundation of China (NSFC) (Grant No. 12374038), Fundamental Research Fund for the Central Universities (Grant No. 2025CDJ-IAISYB-035), and the New Chongqing Youth Innovative Talent Project (Grant No. CSTB2025YITP-QCRCX006), the Major Science and Technology Project of the Chongqing Municipal Education Commission (Grant No. KJZD-M202500504), and the Chongqing Natural Science Foundation-Innovation and Development Joint Fund (Key Program) (Grant No. CSTB2023NSCQ-LZX0138). This work was also supported in part by the NSFC (Grant Nos. 12404045, 52371148, 12547101), the Chongqing Natural Science Foundation (No. CSTB2025NSCQ-GPX1028), the Science and Technology Research Program of Chongqing Municipal Education Commission (No. KJZD-K202500512, No. KJQN-202400553), and the Foundation of Chongqing Normal University (No. 25XLB024 and No. 23XLB015).
 
~~~\\

\textit{Data availability}--The data supporting the findings of this article are publicly available at \cite{Dataava}. 


\begin{thebibliography}{72}%
\makeatletter
\providecommand \@ifxundefined [1]{%
 \@ifx{#1\undefined}
}%
\providecommand \@ifnum [1]{%
 \ifnum #1\expandafter \@firstoftwo
 \else \expandafter \@secondoftwo
 \fi
}%
\providecommand \@ifx [1]{%
 \ifx #1\expandafter \@firstoftwo
 \else \expandafter \@secondoftwo
 \fi
}%
\providecommand \natexlab [1]{#1}%
\providecommand \enquote  [1]{``#1''}%
\providecommand \bibnamefont  [1]{#1}%
\providecommand \bibfnamefont [1]{#1}%
\providecommand \citenamefont [1]{#1}%
\providecommand \href@noop [0]{\@secondoftwo}%
\providecommand \href [0]{\begingroup \@sanitize@url \@href}%
\providecommand \@href[1]{\@@startlink{#1}\@@href}%
\providecommand \@@href[1]{\endgroup#1\@@endlink}%
\providecommand \@sanitize@url [0]{\catcode `\\12\catcode `\$12\catcode
  `\&12\catcode `\#12\catcode `\^12\catcode `\_12\catcode `\%12\relax}%
\providecommand \@@startlink[1]{}%
\providecommand \@@endlink[0]{}%
\providecommand \url  [0]{\begingroup\@sanitize@url \@url }%
\providecommand \@url [1]{\endgroup\@href {#1}{\urlprefix }}%
\providecommand \urlprefix  [0]{URL }%
\providecommand \Eprint [0]{\href }%
\providecommand \doibase [0]{https://doi.org/}%
\providecommand \selectlanguage [0]{\@gobble}%
\providecommand \bibinfo  [0]{\@secondoftwo}%
\providecommand \bibfield  [0]{\@secondoftwo}%
\providecommand \translation [1]{[#1]}%
\providecommand \BibitemOpen [0]{}%
\providecommand \bibitemStop [0]{}%
\providecommand \bibitemNoStop [0]{.\EOS\space}%
\providecommand \EOS [0]{\spacefactor3000\relax}%
\providecommand \BibitemShut  [1]{\csname bibitem#1\endcsname}%
\let\auto@bib@innerbib\@empty
\bibitem [{\citenamefont {Wan}\ \emph {et~al.}(2011)\citenamefont {Wan},
  \citenamefont {Turner}, \citenamefont {Vishwanath},\ and\ \citenamefont
  {Savrasov}}]{wan2011topological}%
  \BibitemOpen
  \bibfield  {author} {\bibinfo {author} {\bibfnamefont {X.}~\bibnamefont
  {Wan}}, \bibinfo {author} {\bibfnamefont {A.~M.}\ \bibnamefont {Turner}},
  \bibinfo {author} {\bibfnamefont {A.}~\bibnamefont {Vishwanath}},\ and\
  \bibinfo {author} {\bibfnamefont {S.~Y.}\ \bibnamefont {Savrasov}},\
  }\bibfield  {title} {\bibinfo {title} {Topological semimetal and fermi-arc
  surface states in the electronic structure of pyrochlore iridates},\ }\href
  {https://doi.org/10.1103/PhysRevB.83.205101} {\bibfield  {journal} {\bibinfo
  {journal} {Phys. Rev. B}\ }\textbf {\bibinfo {volume} {83}},\ \bibinfo
  {pages} {205101} (\bibinfo {year} {2011})}\BibitemShut {NoStop}%
\bibitem [{\citenamefont {Yan}\ and\ \citenamefont
  {Felser}(2017)}]{yan2017topological}%
  \BibitemOpen
  \bibfield  {author} {\bibinfo {author} {\bibfnamefont {B.}~\bibnamefont
  {Yan}}\ and\ \bibinfo {author} {\bibfnamefont {C.}~\bibnamefont {Felser}},\
  }\bibfield  {title} {\bibinfo {title} {Topological materials: Weyl
  semimetals},\ }\href
  {https://doi.org/https://doi.org/10.1146/annurev-conmatphys-031016-025458}
  {\bibfield  {journal} {\bibinfo  {journal} {Annu. Rev. Condens. Matter
  Phys.}\ }\textbf {\bibinfo {volume} {8}},\ \bibinfo {pages} {337} (\bibinfo
  {year} {2017})}\BibitemShut {NoStop}%
\bibitem [{\citenamefont {Armitage}\ \emph {et~al.}(2018)\citenamefont
  {Armitage}, \citenamefont {Mele},\ and\ \citenamefont
  {Vishwanath}}]{armitage2018weyl}%
  \BibitemOpen
  \bibfield  {author} {\bibinfo {author} {\bibfnamefont {N.~P.}\ \bibnamefont
  {Armitage}}, \bibinfo {author} {\bibfnamefont {E.~J.}\ \bibnamefont {Mele}},\
  and\ \bibinfo {author} {\bibfnamefont {A.}~\bibnamefont {Vishwanath}},\
  }\bibfield  {title} {\bibinfo {title} {Weyl and dirac semimetals in
  three-dimensional solids},\ }\href
  {https://doi.org/10.1103/RevModPhys.90.015001} {\bibfield  {journal}
  {\bibinfo  {journal} {Rev. Mod. Phys.}\ }\textbf {\bibinfo {volume} {90}},\
  \bibinfo {pages} {015001} (\bibinfo {year} {2018})}\BibitemShut {NoStop}%
\bibitem [{\citenamefont {Ong}\ and\ \citenamefont
  {Liang}(2021)}]{ong2021experimental}%
  \BibitemOpen
  \bibfield  {author} {\bibinfo {author} {\bibfnamefont {N.}~\bibnamefont
  {Ong}}\ and\ \bibinfo {author} {\bibfnamefont {S.}~\bibnamefont {Liang}},\
  }\bibfield  {title} {\bibinfo {title} {Experimental signatures of the chiral
  anomaly in dirac--weyl semimetals},\ }\href
  {https://doi.org/10.1038/s42254-021-00310-9} {\bibfield  {journal} {\bibinfo
  {journal} {Nat. Rev. Phys.}\ }\textbf {\bibinfo {volume} {3}},\ \bibinfo
  {pages} {394} (\bibinfo {year} {2021})}\BibitemShut {NoStop}%
\bibitem [{\citenamefont {Weng}\ \emph {et~al.}(2015)\citenamefont {Weng},
  \citenamefont {Fang}, \citenamefont {Fang}, \citenamefont {Bernevig},\ and\
  \citenamefont {Dai}}]{PhysRevX.5.011029}%
  \BibitemOpen
  \bibfield  {author} {\bibinfo {author} {\bibfnamefont {H.}~\bibnamefont
  {Weng}}, \bibinfo {author} {\bibfnamefont {C.}~\bibnamefont {Fang}}, \bibinfo
  {author} {\bibfnamefont {Z.}~\bibnamefont {Fang}}, \bibinfo {author}
  {\bibfnamefont {B.~A.}\ \bibnamefont {Bernevig}},\ and\ \bibinfo {author}
  {\bibfnamefont {X.}~\bibnamefont {Dai}},\ }\bibfield  {title} {\bibinfo
  {title} {Weyl semimetal phase in noncentrosymmetric transition-metal
  monophosphides},\ }\href {https://doi.org/10.1103/PhysRevX.5.011029}
  {\bibfield  {journal} {\bibinfo  {journal} {Phys. Rev. X}\ }\textbf {\bibinfo
  {volume} {5}},\ \bibinfo {pages} {011029} (\bibinfo {year}
  {2015})}\BibitemShut {NoStop}%
\bibitem [{\citenamefont {Lv}\ \emph {et~al.}(2015{\natexlab{a}})\citenamefont
  {Lv}, \citenamefont {Weng}, \citenamefont {Fu}, \citenamefont {Wang},
  \citenamefont {Miao}, \citenamefont {Ma}, \citenamefont {Richard},
  \citenamefont {Huang}, \citenamefont {Zhao}, \citenamefont {Chen},
  \citenamefont {Fang}, \citenamefont {Dai}, \citenamefont {Qian},\ and\
  \citenamefont {Ding}}]{lv2015experimental}%
  \BibitemOpen
  \bibfield  {author} {\bibinfo {author} {\bibfnamefont {B.~Q.}\ \bibnamefont
  {Lv}}, \bibinfo {author} {\bibfnamefont {H.~M.}\ \bibnamefont {Weng}},
  \bibinfo {author} {\bibfnamefont {B.~B.}\ \bibnamefont {Fu}}, \bibinfo
  {author} {\bibfnamefont {X.~P.}\ \bibnamefont {Wang}}, \bibinfo {author}
  {\bibfnamefont {H.}~\bibnamefont {Miao}}, \bibinfo {author} {\bibfnamefont
  {J.}~\bibnamefont {Ma}}, \bibinfo {author} {\bibfnamefont {P.}~\bibnamefont
  {Richard}}, \bibinfo {author} {\bibfnamefont {X.~C.}\ \bibnamefont {Huang}},
  \bibinfo {author} {\bibfnamefont {L.~X.}\ \bibnamefont {Zhao}}, \bibinfo
  {author} {\bibfnamefont {G.~F.}\ \bibnamefont {Chen}}, \bibinfo {author}
  {\bibfnamefont {Z.}~\bibnamefont {Fang}}, \bibinfo {author} {\bibfnamefont
  {X.}~\bibnamefont {Dai}}, \bibinfo {author} {\bibfnamefont {T.}~\bibnamefont
  {Qian}},\ and\ \bibinfo {author} {\bibfnamefont {H.}~\bibnamefont {Ding}},\
  }\bibfield  {title} {\bibinfo {title} {Experimental discovery of weyl
  semimetal {TaAs}},\ }\href {https://doi.org/10.1103/PhysRevX.5.031013}
  {\bibfield  {journal} {\bibinfo  {journal} {Phys. Rev. X}\ }\textbf {\bibinfo
  {volume} {5}},\ \bibinfo {pages} {031013} (\bibinfo {year}
  {2015}{\natexlab{a}})}\BibitemShut {NoStop}%
\bibitem [{\citenamefont {Yang}\ \emph {et~al.}(2015)\citenamefont {Yang},
  \citenamefont {Liu}, \citenamefont {Sun}, \citenamefont {Peng}, \citenamefont
  {Yang}, \citenamefont {Zhang}, \citenamefont {Zhou}, \citenamefont {Zhang},
  \citenamefont {Guo}, \citenamefont {Rahn} \emph {et~al.}}]{yang2015weyl}%
  \BibitemOpen
  \bibfield  {author} {\bibinfo {author} {\bibfnamefont {L.}~\bibnamefont
  {Yang}}, \bibinfo {author} {\bibfnamefont {Z.}~\bibnamefont {Liu}}, \bibinfo
  {author} {\bibfnamefont {Y.}~\bibnamefont {Sun}}, \bibinfo {author}
  {\bibfnamefont {H.}~\bibnamefont {Peng}}, \bibinfo {author} {\bibfnamefont
  {H.}~\bibnamefont {Yang}}, \bibinfo {author} {\bibfnamefont {T.}~\bibnamefont
  {Zhang}}, \bibinfo {author} {\bibfnamefont {B.}~\bibnamefont {Zhou}},
  \bibinfo {author} {\bibfnamefont {Y.}~\bibnamefont {Zhang}}, \bibinfo
  {author} {\bibfnamefont {Y.}~\bibnamefont {Guo}}, \bibinfo {author}
  {\bibfnamefont {M.}~\bibnamefont {Rahn}}, \emph {et~al.},\ }\bibfield
  {title} {\bibinfo {title} {Weyl semimetal phase in the non-centrosymmetric
  compound {TaAs}},\ }\href {https://doi.org/10.1038/nphys3425} {\bibfield
  {journal} {\bibinfo  {journal} {Nat. Phys.}\ }\textbf {\bibinfo {volume}
  {11}},\ \bibinfo {pages} {728} (\bibinfo {year} {2015})}\BibitemShut
  {NoStop}%
\bibitem [{\citenamefont {Xu}\ \emph {et~al.}(2015)\citenamefont {Xu},
  \citenamefont {Liu}, \citenamefont {Kushwaha}, \citenamefont {Sankar},
  \citenamefont {Krizan}, \citenamefont {Belopolski}, \citenamefont {Neupane},
  \citenamefont {Bian}, \citenamefont {Alidoust}, \citenamefont {Chang},
  \citenamefont {Jeng}, \citenamefont {Huang}, \citenamefont {Tsai},
  \citenamefont {Lin}, \citenamefont {Shibayev}, \citenamefont {Chou},
  \citenamefont {Cava},\ and\ \citenamefont {Hasan}}]{xu2015observation}%
  \BibitemOpen
  \bibfield  {author} {\bibinfo {author} {\bibfnamefont {S.-Y.}\ \bibnamefont
  {Xu}}, \bibinfo {author} {\bibfnamefont {C.}~\bibnamefont {Liu}}, \bibinfo
  {author} {\bibfnamefont {S.~K.}\ \bibnamefont {Kushwaha}}, \bibinfo {author}
  {\bibfnamefont {R.}~\bibnamefont {Sankar}}, \bibinfo {author} {\bibfnamefont
  {J.~W.}\ \bibnamefont {Krizan}}, \bibinfo {author} {\bibfnamefont
  {I.}~\bibnamefont {Belopolski}}, \bibinfo {author} {\bibfnamefont
  {M.}~\bibnamefont {Neupane}}, \bibinfo {author} {\bibfnamefont
  {G.}~\bibnamefont {Bian}}, \bibinfo {author} {\bibfnamefont {N.}~\bibnamefont
  {Alidoust}}, \bibinfo {author} {\bibfnamefont {T.-R.}\ \bibnamefont {Chang}},
  \bibinfo {author} {\bibfnamefont {H.-T.}\ \bibnamefont {Jeng}}, \bibinfo
  {author} {\bibfnamefont {C.-Y.}\ \bibnamefont {Huang}}, \bibinfo {author}
  {\bibfnamefont {W.-F.}\ \bibnamefont {Tsai}}, \bibinfo {author}
  {\bibfnamefont {H.}~\bibnamefont {Lin}}, \bibinfo {author} {\bibfnamefont
  {P.~P.}\ \bibnamefont {Shibayev}}, \bibinfo {author} {\bibfnamefont {F.-C.}\
  \bibnamefont {Chou}}, \bibinfo {author} {\bibfnamefont {R.~J.}\ \bibnamefont
  {Cava}},\ and\ \bibinfo {author} {\bibfnamefont {M.~Z.}\ \bibnamefont
  {Hasan}},\ }\bibfield  {title} {\bibinfo {title} {Observation of fermi arc
  surface states in a topological metal},\ }\href
  {https://doi.org/10.1126/science.1256742} {\bibfield  {journal} {\bibinfo
  {journal} {Science}\ }\textbf {\bibinfo {volume} {347}},\ \bibinfo {pages}
  {294} (\bibinfo {year} {2015})}\BibitemShut {NoStop}%
\bibitem [{\citenamefont {Sun}\ \emph {et~al.}(2015)\citenamefont {Sun},
  \citenamefont {Wu},\ and\ \citenamefont {Yan}}]{sun2015topological}%
  \BibitemOpen
  \bibfield  {author} {\bibinfo {author} {\bibfnamefont {Y.}~\bibnamefont
  {Sun}}, \bibinfo {author} {\bibfnamefont {S.-C.}\ \bibnamefont {Wu}},\ and\
  \bibinfo {author} {\bibfnamefont {B.}~\bibnamefont {Yan}},\ }\bibfield
  {title} {\bibinfo {title} {Topological surface states and fermi arcs of the
  noncentrosymmetric weyl semimetals {TaAs}, {TaP}, {NbAs}, and {NbP}},\ }\href
  {https://doi.org/10.1103/PhysRevB.92.115428} {\bibfield  {journal} {\bibinfo
  {journal} {Phys. Rev. B}\ }\textbf {\bibinfo {volume} {92}},\ \bibinfo
  {pages} {115428} (\bibinfo {year} {2015})}\BibitemShut {NoStop}%
\bibitem [{\citenamefont {Shekhar}\ \emph
  {et~al.}(2015{\natexlab{a}})\citenamefont {Shekhar}, \citenamefont {Nayak},
  \citenamefont {Sun}, \citenamefont {Schmidt}, \citenamefont {Nicklas},
  \citenamefont {Leermakers}, \citenamefont {Zeitler}, \citenamefont
  {Skourski}, \citenamefont {Wosnitza}, \citenamefont {Liu} \emph
  {et~al.}}]{shekhar2015extremely}%
  \BibitemOpen
  \bibfield  {author} {\bibinfo {author} {\bibfnamefont {C.}~\bibnamefont
  {Shekhar}}, \bibinfo {author} {\bibfnamefont {A.~K.}\ \bibnamefont {Nayak}},
  \bibinfo {author} {\bibfnamefont {Y.}~\bibnamefont {Sun}}, \bibinfo {author}
  {\bibfnamefont {M.}~\bibnamefont {Schmidt}}, \bibinfo {author} {\bibfnamefont
  {M.}~\bibnamefont {Nicklas}}, \bibinfo {author} {\bibfnamefont
  {I.}~\bibnamefont {Leermakers}}, \bibinfo {author} {\bibfnamefont
  {U.}~\bibnamefont {Zeitler}}, \bibinfo {author} {\bibfnamefont
  {Y.}~\bibnamefont {Skourski}}, \bibinfo {author} {\bibfnamefont
  {J.}~\bibnamefont {Wosnitza}}, \bibinfo {author} {\bibfnamefont
  {Z.}~\bibnamefont {Liu}}, \emph {et~al.},\ }\bibfield  {title} {\bibinfo
  {title} {Extremely large magnetoresistance and ultrahigh mobility in the
  topological weyl semimetal candidate {NbP}},\ }\href
  {https://doi.org/10.1038/nphys3372} {\bibfield  {journal} {\bibinfo
  {journal} {Nat. Phys.}\ }\textbf {\bibinfo {volume} {11}},\ \bibinfo {pages}
  {645} (\bibinfo {year} {2015}{\natexlab{a}})}\BibitemShut {NoStop}%
\bibitem [{\citenamefont {Shekhar}\ \emph
  {et~al.}(2015{\natexlab{b}})\citenamefont {Shekhar}, \citenamefont {Arnold},
  \citenamefont {Wu}, \citenamefont {Sun}, \citenamefont {Schmidt},
  \citenamefont {Kumar}, \citenamefont {Grushin}, \citenamefont {Bardarson},
  \citenamefont {dos Reis}, \citenamefont {Naumann} \emph
  {et~al.}}]{shekhar2015large}%
  \BibitemOpen
  \bibfield  {author} {\bibinfo {author} {\bibfnamefont {C.}~\bibnamefont
  {Shekhar}}, \bibinfo {author} {\bibfnamefont {F.}~\bibnamefont {Arnold}},
  \bibinfo {author} {\bibfnamefont {S.-C.}\ \bibnamefont {Wu}}, \bibinfo
  {author} {\bibfnamefont {Y.}~\bibnamefont {Sun}}, \bibinfo {author}
  {\bibfnamefont {M.}~\bibnamefont {Schmidt}}, \bibinfo {author} {\bibfnamefont
  {N.}~\bibnamefont {Kumar}}, \bibinfo {author} {\bibfnamefont {A.~G.}\
  \bibnamefont {Grushin}}, \bibinfo {author} {\bibfnamefont {J.~H.}\
  \bibnamefont {Bardarson}}, \bibinfo {author} {\bibfnamefont {R.~D.}\
  \bibnamefont {dos Reis}}, \bibinfo {author} {\bibfnamefont {M.}~\bibnamefont
  {Naumann}}, \emph {et~al.},\ }\bibfield  {title} {\bibinfo {title} {Large and
  unsaturated negative magnetoresistance induced by the chiral anomaly in the
  weyl semimetal {TaP}},\ }\href {https://doi.org/10.48550/arXiv.1506.06577}
  {\bibfield  {journal} {\bibinfo  {journal} {Preprint at http://arxiv.
  org/abs/1506.06577}\ } (\bibinfo {year} {2015}{\natexlab{b}})}\BibitemShut
  {NoStop}%
\bibitem [{\citenamefont {Du}\ \emph {et~al.}(2016)\citenamefont {Du},
  \citenamefont {Wang}, \citenamefont {Chen}, \citenamefont {Mao},
  \citenamefont {Khan}, \citenamefont {Xu}, \citenamefont {Zhou}, \citenamefont
  {Zhang}, \citenamefont {Yang}, \citenamefont {Chen} \emph
  {et~al.}}]{du2016large}%
  \BibitemOpen
  \bibfield  {author} {\bibinfo {author} {\bibfnamefont {J.}~\bibnamefont
  {Du}}, \bibinfo {author} {\bibfnamefont {H.}~\bibnamefont {Wang}}, \bibinfo
  {author} {\bibfnamefont {Q.}~\bibnamefont {Chen}}, \bibinfo {author}
  {\bibfnamefont {Q.}~\bibnamefont {Mao}}, \bibinfo {author} {\bibfnamefont
  {R.}~\bibnamefont {Khan}}, \bibinfo {author} {\bibfnamefont {B.}~\bibnamefont
  {Xu}}, \bibinfo {author} {\bibfnamefont {Y.}~\bibnamefont {Zhou}}, \bibinfo
  {author} {\bibfnamefont {Y.}~\bibnamefont {Zhang}}, \bibinfo {author}
  {\bibfnamefont {J.}~\bibnamefont {Yang}}, \bibinfo {author} {\bibfnamefont
  {B.}~\bibnamefont {Chen}}, \emph {et~al.},\ }\bibfield  {title} {\bibinfo
  {title} {Large unsaturated positive and negative magnetoresistance in weyl
  semimetal {TaP}},\ }\href {https://doi.org/10.1007/s11433-016-5798-4}
  {\bibfield  {journal} {\bibinfo  {journal} {Sci. China Phys. Mech. Astron.}\
  }\textbf {\bibinfo {volume} {59}},\ \bibinfo {pages} {657406} (\bibinfo
  {year} {2016})}\BibitemShut {NoStop}%
\bibitem [{\citenamefont {Zhang}\ \emph {et~al.}(2019)\citenamefont {Zhang},
  \citenamefont {Wang}, \citenamefont {Yuan}, \citenamefont {Xu}, \citenamefont
  {Wang}, \citenamefont {Lee}, \citenamefont {Pi}, \citenamefont {Xi},
  \citenamefont {Lin}, \citenamefont {Harrison} \emph {et~al.}}]{zhang2019non}%
  \BibitemOpen
  \bibfield  {author} {\bibinfo {author} {\bibfnamefont {C.-L.}\ \bibnamefont
  {Zhang}}, \bibinfo {author} {\bibfnamefont {C.}~\bibnamefont {Wang}},
  \bibinfo {author} {\bibfnamefont {Z.}~\bibnamefont {Yuan}}, \bibinfo {author}
  {\bibfnamefont {X.}~\bibnamefont {Xu}}, \bibinfo {author} {\bibfnamefont
  {G.}~\bibnamefont {Wang}}, \bibinfo {author} {\bibfnamefont {C.-C.}\
  \bibnamefont {Lee}}, \bibinfo {author} {\bibfnamefont {L.}~\bibnamefont
  {Pi}}, \bibinfo {author} {\bibfnamefont {C.}~\bibnamefont {Xi}}, \bibinfo
  {author} {\bibfnamefont {H.}~\bibnamefont {Lin}}, \bibinfo {author}
  {\bibfnamefont {N.}~\bibnamefont {Harrison}}, \emph {et~al.},\ }\bibfield
  {title} {\bibinfo {title} {Non-saturating quantum magnetization in weyl
  semimetal {TaAs}},\ }\href {https://doi.org/10.1038/s41467-019-09012-4}
  {\bibfield  {journal} {\bibinfo  {journal} {Nat. Commun.}\ }\textbf {\bibinfo
  {volume} {10}},\ \bibinfo {pages} {1028} (\bibinfo {year}
  {2019})}\BibitemShut {NoStop}%
\bibitem [{\citenamefont {Arnold}\ \emph {et~al.}(2016)\citenamefont {Arnold},
  \citenamefont {Shekhar}, \citenamefont {Wu}, \citenamefont {Sun},
  \citenamefont {Dos~Reis}, \citenamefont {Kumar}, \citenamefont {Naumann},
  \citenamefont {Ajeesh}, \citenamefont {Schmidt}, \citenamefont {Grushin}
  \emph {et~al.}}]{arnold2016negative}%
  \BibitemOpen
  \bibfield  {author} {\bibinfo {author} {\bibfnamefont {F.}~\bibnamefont
  {Arnold}}, \bibinfo {author} {\bibfnamefont {C.}~\bibnamefont {Shekhar}},
  \bibinfo {author} {\bibfnamefont {S.-C.}\ \bibnamefont {Wu}}, \bibinfo
  {author} {\bibfnamefont {Y.}~\bibnamefont {Sun}}, \bibinfo {author}
  {\bibfnamefont {R.~D.}\ \bibnamefont {Dos~Reis}}, \bibinfo {author}
  {\bibfnamefont {N.}~\bibnamefont {Kumar}}, \bibinfo {author} {\bibfnamefont
  {M.}~\bibnamefont {Naumann}}, \bibinfo {author} {\bibfnamefont {M.~O.}\
  \bibnamefont {Ajeesh}}, \bibinfo {author} {\bibfnamefont {M.}~\bibnamefont
  {Schmidt}}, \bibinfo {author} {\bibfnamefont {A.~G.}\ \bibnamefont
  {Grushin}}, \emph {et~al.},\ }\bibfield  {title} {\bibinfo {title} {Negative
  magnetoresistance without well-defined chirality in the weyl semimetal
  {TaP}},\ }\href {https://doi.org/10.1038/ncomms11615} {\bibfield  {journal}
  {\bibinfo  {journal} {Nat. Commun.}\ }\textbf {\bibinfo {volume} {7}},\
  \bibinfo {pages} {11615} (\bibinfo {year} {2016})}\BibitemShut {NoStop}%
\bibitem [{\citenamefont {Huang}\ \emph {et~al.}(2015)\citenamefont {Huang},
  \citenamefont {Zhao}, \citenamefont {Long}, \citenamefont {Wang},
  \citenamefont {Chen}, \citenamefont {Yang}, \citenamefont {Liang},
  \citenamefont {Xue}, \citenamefont {Weng}, \citenamefont {Fang},
  \citenamefont {Dai},\ and\ \citenamefont {Chen}}]{PhysRevX.5.031023}%
  \BibitemOpen
  \bibfield  {author} {\bibinfo {author} {\bibfnamefont {X.}~\bibnamefont
  {Huang}}, \bibinfo {author} {\bibfnamefont {L.}~\bibnamefont {Zhao}},
  \bibinfo {author} {\bibfnamefont {Y.}~\bibnamefont {Long}}, \bibinfo {author}
  {\bibfnamefont {P.}~\bibnamefont {Wang}}, \bibinfo {author} {\bibfnamefont
  {D.}~\bibnamefont {Chen}}, \bibinfo {author} {\bibfnamefont {Z.}~\bibnamefont
  {Yang}}, \bibinfo {author} {\bibfnamefont {H.}~\bibnamefont {Liang}},
  \bibinfo {author} {\bibfnamefont {M.}~\bibnamefont {Xue}}, \bibinfo {author}
  {\bibfnamefont {H.}~\bibnamefont {Weng}}, \bibinfo {author} {\bibfnamefont
  {Z.}~\bibnamefont {Fang}}, \bibinfo {author} {\bibfnamefont {X.}~\bibnamefont
  {Dai}},\ and\ \bibinfo {author} {\bibfnamefont {G.}~\bibnamefont {Chen}},\
  }\bibfield  {title} {\bibinfo {title} {Observation of the
  chiral-anomaly-induced negative magnetoresistance in 3d weyl semimetal
  {TaAs}},\ }\href {https://doi.org/10.1103/PhysRevX.5.031023} {\bibfield
  {journal} {\bibinfo  {journal} {Phys. Rev. X}\ }\textbf {\bibinfo {volume}
  {5}},\ \bibinfo {pages} {031023} (\bibinfo {year} {2015})}\BibitemShut
  {NoStop}%
\bibitem [{\citenamefont {Li}\ \emph {et~al.}(2017)\citenamefont {Li},
  \citenamefont {Wang}, \citenamefont {Li}, \citenamefont {Yang}, \citenamefont
  {Shen}, \citenamefont {Sheng}, \citenamefont {Li}, \citenamefont {Lu},
  \citenamefont {Zheng},\ and\ \citenamefont {Xu}}]{li2017negative}%
  \BibitemOpen
  \bibfield  {author} {\bibinfo {author} {\bibfnamefont {Y.}~\bibnamefont
  {Li}}, \bibinfo {author} {\bibfnamefont {Z.}~\bibnamefont {Wang}}, \bibinfo
  {author} {\bibfnamefont {P.}~\bibnamefont {Li}}, \bibinfo {author}
  {\bibfnamefont {X.}~\bibnamefont {Yang}}, \bibinfo {author} {\bibfnamefont
  {Z.}~\bibnamefont {Shen}}, \bibinfo {author} {\bibfnamefont {F.}~\bibnamefont
  {Sheng}}, \bibinfo {author} {\bibfnamefont {X.}~\bibnamefont {Li}}, \bibinfo
  {author} {\bibfnamefont {Y.}~\bibnamefont {Lu}}, \bibinfo {author}
  {\bibfnamefont {Y.}~\bibnamefont {Zheng}},\ and\ \bibinfo {author}
  {\bibfnamefont {Z.-A.}\ \bibnamefont {Xu}},\ }\bibfield  {title} {\bibinfo
  {title} {Negative magnetoresistance in weyl semimetals {NbAs} and {NbP}:
  Intrinsic chiral anomaly and extrinsic effects},\ }\href
  {https://doi.org/10.1007/s11467-016-0636-8} {\bibfield  {journal} {\bibinfo
  {journal} {Front. Phys.}\ }\textbf {\bibinfo {volume} {12}},\ \bibinfo
  {pages} {127205} (\bibinfo {year} {2017})}\BibitemShut {NoStop}%
\bibitem [{\citenamefont {Ashby}\ and\ \citenamefont
  {Carbotte}(2014)}]{ashby2014chiral}%
  \BibitemOpen
  \bibfield  {author} {\bibinfo {author} {\bibfnamefont {P.~E.~C.}\
  \bibnamefont {Ashby}}\ and\ \bibinfo {author} {\bibfnamefont {J.~P.}\
  \bibnamefont {Carbotte}},\ }\bibfield  {title} {\bibinfo {title} {Chiral
  anomaly and optical absorption in weyl semimetals},\ }\href
  {https://doi.org/10.1103/PhysRevB.89.245121} {\bibfield  {journal} {\bibinfo
  {journal} {Phys. Rev. B}\ }\textbf {\bibinfo {volume} {89}},\ \bibinfo
  {pages} {245121} (\bibinfo {year} {2014})}\BibitemShut {NoStop}%
\bibitem [{\citenamefont {Jia}\ \emph {et~al.}(2016)\citenamefont {Jia},
  \citenamefont {Xu},\ and\ \citenamefont {Hasan}}]{jia2016weyl}%
  \BibitemOpen
  \bibfield  {author} {\bibinfo {author} {\bibfnamefont {S.}~\bibnamefont
  {Jia}}, \bibinfo {author} {\bibfnamefont {S.-Y.}\ \bibnamefont {Xu}},\ and\
  \bibinfo {author} {\bibfnamefont {M.~Z.}\ \bibnamefont {Hasan}},\ }\bibfield
  {title} {\bibinfo {title} {Weyl semimetals, fermi arcs and chiral
  anomalies},\ }\href {https://doi.org/10.1038/nmat4787} {\bibfield  {journal}
  {\bibinfo  {journal} {Nat. Mater.}\ }\textbf {\bibinfo {volume} {15}},\
  \bibinfo {pages} {1140} (\bibinfo {year} {2016})}\BibitemShut {NoStop}%
\bibitem [{\citenamefont {Yuan}\ \emph {et~al.}(2020)\citenamefont {Yuan},
  \citenamefont {Zhang}, \citenamefont {Zhang}, \citenamefont {Yan},
  \citenamefont {Lyu}, \citenamefont {Zhang}, \citenamefont {Li}, \citenamefont
  {Song}, \citenamefont {Zhao}, \citenamefont {Leng} \emph
  {et~al.}}]{yuan2020discovery}%
  \BibitemOpen
  \bibfield  {author} {\bibinfo {author} {\bibfnamefont {X.}~\bibnamefont
  {Yuan}}, \bibinfo {author} {\bibfnamefont {C.}~\bibnamefont {Zhang}},
  \bibinfo {author} {\bibfnamefont {Y.}~\bibnamefont {Zhang}}, \bibinfo
  {author} {\bibfnamefont {Z.}~\bibnamefont {Yan}}, \bibinfo {author}
  {\bibfnamefont {T.}~\bibnamefont {Lyu}}, \bibinfo {author} {\bibfnamefont
  {M.}~\bibnamefont {Zhang}}, \bibinfo {author} {\bibfnamefont
  {Z.}~\bibnamefont {Li}}, \bibinfo {author} {\bibfnamefont {C.}~\bibnamefont
  {Song}}, \bibinfo {author} {\bibfnamefont {M.}~\bibnamefont {Zhao}}, \bibinfo
  {author} {\bibfnamefont {P.}~\bibnamefont {Leng}}, \emph {et~al.},\
  }\bibfield  {title} {\bibinfo {title} {The discovery of dynamic chiral
  anomaly in a weyl semimetal {NbAs}},\ }\href
  {https://doi.org/10.1038/s41467-020-14749-4} {\bibfield  {journal} {\bibinfo
  {journal} {Nat. Commun.}\ }\textbf {\bibinfo {volume} {11}},\ \bibinfo
  {pages} {1259} (\bibinfo {year} {2020})}\BibitemShut {NoStop}%
\bibitem [{\citenamefont {Xu}\ \emph {et~al.}(2016)\citenamefont {Xu},
  \citenamefont {Dai}, \citenamefont {Zhao}, \citenamefont {Wang},
  \citenamefont {Yang}, \citenamefont {Zhang}, \citenamefont {Liu},
  \citenamefont {Xiao}, \citenamefont {Chen}, \citenamefont {Taylor},
  \citenamefont {Yarotski}, \citenamefont {Prasankumar},\ and\ \citenamefont
  {Qiu}}]{xu2016optical}%
  \BibitemOpen
  \bibfield  {author} {\bibinfo {author} {\bibfnamefont {B.}~\bibnamefont
  {Xu}}, \bibinfo {author} {\bibfnamefont {Y.~M.}\ \bibnamefont {Dai}},
  \bibinfo {author} {\bibfnamefont {L.~X.}\ \bibnamefont {Zhao}}, \bibinfo
  {author} {\bibfnamefont {K.}~\bibnamefont {Wang}}, \bibinfo {author}
  {\bibfnamefont {R.}~\bibnamefont {Yang}}, \bibinfo {author} {\bibfnamefont
  {W.}~\bibnamefont {Zhang}}, \bibinfo {author} {\bibfnamefont {J.~Y.}\
  \bibnamefont {Liu}}, \bibinfo {author} {\bibfnamefont {H.}~\bibnamefont
  {Xiao}}, \bibinfo {author} {\bibfnamefont {G.~F.}\ \bibnamefont {Chen}},
  \bibinfo {author} {\bibfnamefont {A.~J.}\ \bibnamefont {Taylor}}, \bibinfo
  {author} {\bibfnamefont {D.~A.}\ \bibnamefont {Yarotski}}, \bibinfo {author}
  {\bibfnamefont {R.~P.}\ \bibnamefont {Prasankumar}},\ and\ \bibinfo {author}
  {\bibfnamefont {X.~G.}\ \bibnamefont {Qiu}},\ }\bibfield  {title} {\bibinfo
  {title} {Optical spectroscopy of the weyl semimetal {TaAs}},\ }\href
  {https://doi.org/10.1103/PhysRevB.93.121110} {\bibfield  {journal} {\bibinfo
  {journal} {Phys. Rev. B}\ }\textbf {\bibinfo {volume} {93}},\ \bibinfo
  {pages} {121110} (\bibinfo {year} {2016})}\BibitemShut {NoStop}%
\bibitem [{\citenamefont {Ding}\ \emph {et~al.}(2024)\citenamefont {Ding},
  \citenamefont {Jin}, \citenamefont {Chang}, \citenamefont {Li}, \citenamefont
  {Zhou}, \citenamefont {Yang},\ and\ \citenamefont
  {Wang}}]{PhysRevB.110.054304}%
  \BibitemOpen
  \bibfield  {author} {\bibinfo {author} {\bibfnamefont {X.}~\bibnamefont
  {Ding}}, \bibinfo {author} {\bibfnamefont {X.}~\bibnamefont {Jin}}, \bibinfo
  {author} {\bibfnamefont {Z.}~\bibnamefont {Chang}}, \bibinfo {author}
  {\bibfnamefont {D.}~\bibnamefont {Li}}, \bibinfo {author} {\bibfnamefont
  {X.}~\bibnamefont {Zhou}}, \bibinfo {author} {\bibfnamefont {X.}~\bibnamefont
  {Yang}},\ and\ \bibinfo {author} {\bibfnamefont {R.}~\bibnamefont {Wang}},\
  }\bibfield  {title} {\bibinfo {title} {Anharmonicity-induced phonon hardening
  and anomalous thermal transport in {ScZn}},\ }\href
  {https://doi.org/10.1103/PhysRevB.110.054304} {\bibfield  {journal} {\bibinfo
   {journal} {Phys. Rev. B}\ }\textbf {\bibinfo {volume} {110}},\ \bibinfo
  {pages} {054304} (\bibinfo {year} {2024})}\BibitemShut {NoStop}%
\bibitem [{\citenamefont {Kundu}\ \emph {et~al.}(2021)\citenamefont {Kundu},
  \citenamefont {Yang}, \citenamefont {Ma}, \citenamefont {Feng}, \citenamefont
  {Carrete}, \citenamefont {Ruan}, \citenamefont {Madsen},\ and\ \citenamefont
  {Li}}]{kundu2021ultrahigh}%
  \BibitemOpen
  \bibfield  {author} {\bibinfo {author} {\bibfnamefont {A.}~\bibnamefont
  {Kundu}}, \bibinfo {author} {\bibfnamefont {X.}~\bibnamefont {Yang}},
  \bibinfo {author} {\bibfnamefont {J.}~\bibnamefont {Ma}}, \bibinfo {author}
  {\bibfnamefont {T.}~\bibnamefont {Feng}}, \bibinfo {author} {\bibfnamefont
  {J.}~\bibnamefont {Carrete}}, \bibinfo {author} {\bibfnamefont
  {X.}~\bibnamefont {Ruan}}, \bibinfo {author} {\bibfnamefont {G.~K.~H.}\
  \bibnamefont {Madsen}},\ and\ \bibinfo {author} {\bibfnamefont
  {W.}~\bibnamefont {Li}},\ }\bibfield  {title} {\bibinfo {title} {Ultrahigh
  thermal conductivity of $\ensuremath{\theta}$-phase tantalum nitride},\
  }\href {https://doi.org/10.1103/PhysRevLett.126.115901} {\bibfield  {journal}
  {\bibinfo  {journal} {Phys. Rev. Lett.}\ }\textbf {\bibinfo {volume} {126}},\
  \bibinfo {pages} {115901} (\bibinfo {year} {2021})}\BibitemShut {NoStop}%
\bibitem [{\citenamefont {Jin}\ \emph {et~al.}(2025)\citenamefont {Jin},
  \citenamefont {Zhang}, \citenamefont {Li}, \citenamefont {Cheng},
  \citenamefont {Wang}, \citenamefont {Lv}, \citenamefont {Zhou}, \citenamefont
  {Wang}, \citenamefont {Ding}, \citenamefont {Yu},\ and\ \citenamefont
  {Yang}}]{jin2025anharmonicity}%
  \BibitemOpen
  \bibfield  {author} {\bibinfo {author} {\bibfnamefont {X.}~\bibnamefont
  {Jin}}, \bibinfo {author} {\bibfnamefont {Q.}~\bibnamefont {Zhang}}, \bibinfo
  {author} {\bibfnamefont {D.}~\bibnamefont {Li}}, \bibinfo {author}
  {\bibfnamefont {Z.}~\bibnamefont {Cheng}}, \bibinfo {author} {\bibfnamefont
  {J.}~\bibnamefont {Wang}}, \bibinfo {author} {\bibfnamefont {X.}~\bibnamefont
  {Lv}}, \bibinfo {author} {\bibfnamefont {X.}~\bibnamefont {Zhou}}, \bibinfo
  {author} {\bibfnamefont {R.}~\bibnamefont {Wang}}, \bibinfo {author}
  {\bibfnamefont {X.}~\bibnamefont {Ding}}, \bibinfo {author} {\bibfnamefont
  {P.}~\bibnamefont {Yu}},\ and\ \bibinfo {author} {\bibfnamefont
  {X.}~\bibnamefont {Yang}},\ }\bibfield  {title} {\bibinfo {title}
  {Anharmonicity-driven avoided phonon crossing and anomalous thermal transport
  in the nodal-line semimetal {ZrSiS}},\ }\href
  {https://doi.org/10.1103/zzb2-tqfg} {\bibfield  {journal} {\bibinfo
  {journal} {Phys. Rev. B}\ }\textbf {\bibinfo {volume} {112}},\ \bibinfo
  {pages} {184311} (\bibinfo {year} {2025})}\BibitemShut {NoStop}%
\bibitem [{\citenamefont {Chester}\ and\ \citenamefont
  {Thellung}(1961)}]{chester1961law}%
  \BibitemOpen
  \bibfield  {author} {\bibinfo {author} {\bibfnamefont {G.~V.}\ \bibnamefont
  {Chester}}\ and\ \bibinfo {author} {\bibfnamefont {A.}~\bibnamefont
  {Thellung}},\ }\bibfield  {title} {\bibinfo {title} {The law of wiedemann and
  franz},\ }\href {https://doi.org/10.1088/0370-1328/77/5/309} {\bibfield
  {journal} {\bibinfo  {journal} {Proc. Phys. Soc}\ }\textbf {\bibinfo {volume}
  {77}},\ \bibinfo {pages} {1005} (\bibinfo {year} {1961})}\BibitemShut
  {NoStop}%
\bibitem [{\citenamefont {Makinson}(1938)}]{makinson1938thermal}%
  \BibitemOpen
  \bibfield  {author} {\bibinfo {author} {\bibfnamefont {R.}~\bibnamefont
  {Makinson}},\ }\bibfield  {title} {\bibinfo {title} {The thermal conductivity
  of metals},\ }in\ \href {https://doi.org/10.1017/S0305004100020442} {\emph
  {\bibinfo {booktitle} {Mathematical Proceedings of the Cambridge
  Philosophical Society}}},\ Vol.~\bibinfo {volume} {34}\ (\bibinfo
  {organization} {Cambridge University Press},\ \bibinfo {year} {1938})\ pp.\
  \bibinfo {pages} {474--497}\BibitemShut {NoStop}%
\bibitem [{\citenamefont {Klemens}\ and\ \citenamefont
  {Williams}(1986)}]{klemens1986thermal}%
  \BibitemOpen
  \bibfield  {author} {\bibinfo {author} {\bibfnamefont {P.~G.}\ \bibnamefont
  {Klemens}}\ and\ \bibinfo {author} {\bibfnamefont {R.~K.}\ \bibnamefont
  {Williams}},\ }\bibfield  {title} {\bibinfo {title} {Thermal conductivity of
  metals and alloys},\ }\href {https://doi.org/10.1179/imtr.1986.31.1.197}
  {\bibfield  {journal} {\bibinfo  {journal} {Int. Mater. Rev.}\ }\textbf
  {\bibinfo {volume} {31}},\ \bibinfo {pages} {197} (\bibinfo {year}
  {1986})}\BibitemShut {NoStop}%
\bibitem [{\citenamefont {Chen}\ \emph {et~al.}(2019)\citenamefont {Chen},
  \citenamefont {Ma},\ and\ \citenamefont {Li}}]{PhysRevB.99.020305}%
  \BibitemOpen
  \bibfield  {author} {\bibinfo {author} {\bibfnamefont {Y.}~\bibnamefont
  {Chen}}, \bibinfo {author} {\bibfnamefont {J.}~\bibnamefont {Ma}},\ and\
  \bibinfo {author} {\bibfnamefont {W.}~\bibnamefont {Li}},\ }\bibfield
  {title} {\bibinfo {title} {Understanding the thermal conductivity and lorenz
  number in tungsten from first principles},\ }\href
  {https://doi.org/10.1103/PhysRevB.99.020305} {\bibfield  {journal} {\bibinfo
  {journal} {Phys. Rev. B}\ }\textbf {\bibinfo {volume} {99}},\ \bibinfo
  {pages} {020305} (\bibinfo {year} {2019})}\BibitemShut {NoStop}%
\bibitem [{\citenamefont {Chen}\ \emph {et~al.}(2024)\citenamefont {Chen},
  \citenamefont {Pang}, \citenamefont {Meng},\ and\ \citenamefont
  {Li}}]{PhysRevB.109.L220302}%
  \BibitemOpen
  \bibfield  {author} {\bibinfo {author} {\bibfnamefont {Y.}~\bibnamefont
  {Chen}}, \bibinfo {author} {\bibfnamefont {G.}~\bibnamefont {Pang}}, \bibinfo
  {author} {\bibfnamefont {F.}~\bibnamefont {Meng}},\ and\ \bibinfo {author}
  {\bibfnamefont {W.}~\bibnamefont {Li}},\ }\bibfield  {title} {\bibinfo
  {title} {Origin of the high lattice thermal conductivity of beryllium among
  the elemental metals},\ }\href {https://doi.org/10.1103/PhysRevB.109.L220302}
  {\bibfield  {journal} {\bibinfo  {journal} {Phys. Rev. B}\ }\textbf {\bibinfo
  {volume} {109}},\ \bibinfo {pages} {L220302} (\bibinfo {year}
  {2024})}\BibitemShut {NoStop}%
\bibitem [{\citenamefont {Tong}\ \emph {et~al.}(2019)\citenamefont {Tong},
  \citenamefont {Li}, \citenamefont {Ruan},\ and\ \citenamefont
  {Bao}}]{PhysRevB.100.144306}%
  \BibitemOpen
  \bibfield  {author} {\bibinfo {author} {\bibfnamefont {Z.}~\bibnamefont
  {Tong}}, \bibinfo {author} {\bibfnamefont {S.}~\bibnamefont {Li}}, \bibinfo
  {author} {\bibfnamefont {X.}~\bibnamefont {Ruan}},\ and\ \bibinfo {author}
  {\bibfnamefont {H.}~\bibnamefont {Bao}},\ }\bibfield  {title} {\bibinfo
  {title} {Comprehensive first-principles analysis of phonon thermal
  conductivity and electron-phonon coupling in different metals},\ }\href
  {https://doi.org/10.1103/PhysRevB.100.144306} {\bibfield  {journal} {\bibinfo
   {journal} {Phys. Rev. B}\ }\textbf {\bibinfo {volume} {100}},\ \bibinfo
  {pages} {144306} (\bibinfo {year} {2019})}\BibitemShut {NoStop}%
\bibitem [{\citenamefont {Li}\ \emph {et~al.}(2026)\citenamefont {Li},
  \citenamefont {Su}, \citenamefont {Qin}, \citenamefont {Alatas},
  \citenamefont {Kunz}, \citenamefont {Yamada}, \citenamefont {Kelly},
  \citenamefont {Upton}, \citenamefont {Gironda}, \citenamefont {Zhao},
  \citenamefont {Kalkan}, \citenamefont {Yang}, \citenamefont {Aoki},\ and\
  \citenamefont {Hu}}]{doi:10.1126/science.aeb1142}%
  \BibitemOpen
  \bibfield  {author} {\bibinfo {author} {\bibfnamefont {S.}~\bibnamefont
  {Li}}, \bibinfo {author} {\bibfnamefont {C.}~\bibnamefont {Su}}, \bibinfo
  {author} {\bibfnamefont {Z.}~\bibnamefont {Qin}}, \bibinfo {author}
  {\bibfnamefont {A.}~\bibnamefont {Alatas}}, \bibinfo {author} {\bibfnamefont
  {M.}~\bibnamefont {Kunz}}, \bibinfo {author} {\bibfnamefont {T.}~\bibnamefont
  {Yamada}}, \bibinfo {author} {\bibfnamefont {S.~D.}\ \bibnamefont {Kelly}},
  \bibinfo {author} {\bibfnamefont {M.~H.}\ \bibnamefont {Upton}}, \bibinfo
  {author} {\bibfnamefont {A.}~\bibnamefont {Gironda}}, \bibinfo {author}
  {\bibfnamefont {J.}~\bibnamefont {Zhao}}, \bibinfo {author} {\bibfnamefont
  {B.}~\bibnamefont {Kalkan}}, \bibinfo {author} {\bibfnamefont
  {W.}~\bibnamefont {Yang}}, \bibinfo {author} {\bibfnamefont {T.}~\bibnamefont
  {Aoki}},\ and\ \bibinfo {author} {\bibfnamefont {Y.}~\bibnamefont {Hu}},\
  }\bibfield  {title} {\bibinfo {title} {Metallic $\theta$-phase tantalum
  nitride has a thermal conductivity triple that of copper},\ }\href
  {https://doi.org/10.1126/science.aeb1142} {\bibfield  {journal} {\bibinfo
  {journal} {Science}\ }\textbf {\bibinfo {volume} {391}},\ \bibinfo {pages}
  {707} (\bibinfo {year} {2026})}\BibitemShut {NoStop}%
\bibitem [{\citenamefont {Liu}\ \emph {et~al.}(2026)\citenamefont {Liu},
  \citenamefont {Zhou}, \citenamefont {Pang}, \citenamefont {Gu}, \citenamefont
  {Song}, \citenamefont {Chen}, \citenamefont {Carrete}, \citenamefont {Fang},
  \citenamefont {Wang}, \citenamefont {Li},\ and\ \citenamefont
  {Sun}}]{10.1093/nsr/nwag106}%
  \BibitemOpen
  \bibfield  {author} {\bibinfo {author} {\bibfnamefont {Y.}~\bibnamefont
  {Liu}}, \bibinfo {author} {\bibfnamefont {X.}~\bibnamefont {Zhou}}, \bibinfo
  {author} {\bibfnamefont {G.}~\bibnamefont {Pang}}, \bibinfo {author}
  {\bibfnamefont {C.}~\bibnamefont {Gu}}, \bibinfo {author} {\bibfnamefont
  {G.}~\bibnamefont {Song}}, \bibinfo {author} {\bibfnamefont {J.}~\bibnamefont
  {Chen}}, \bibinfo {author} {\bibfnamefont {J.}~\bibnamefont {Carrete}},
  \bibinfo {author} {\bibfnamefont {L.}~\bibnamefont {Fang}}, \bibinfo {author}
  {\bibfnamefont {S.}~\bibnamefont {Wang}}, \bibinfo {author} {\bibfnamefont
  {W.}~\bibnamefont {Li}},\ and\ \bibinfo {author} {\bibfnamefont
  {B.}~\bibnamefont {Sun}},\ }\bibfield  {title} {\bibinfo {title} {High
  thermal conductivity in metallic {$\theta$-TaN} single crystals},\ }\href
  {https://doi.org/10.1093/nsr/nwag106} {\bibfield  {journal} {\bibinfo
  {journal} {Nat. Sci. Rev.}\ }\textbf {\bibinfo {volume} {13}},\ \bibinfo
  {pages} {nwag106} (\bibinfo {year} {2026})}\BibitemShut {NoStop}%
\bibitem [{\citenamefont {Ding}\ \emph {et~al.}(2025)\citenamefont {Ding},
  \citenamefont {Jin}, \citenamefont {Li}, \citenamefont {Fan}, \citenamefont
  {Zhou}, \citenamefont {Lv}, \citenamefont {Yang}, \citenamefont {Cheng},\
  and\ \citenamefont {Wang}}]{ding2025concurrent}%
  \BibitemOpen
  \bibfield  {author} {\bibinfo {author} {\bibfnamefont {X.}~\bibnamefont
  {Ding}}, \bibinfo {author} {\bibfnamefont {X.}~\bibnamefont {Jin}}, \bibinfo
  {author} {\bibfnamefont {D.}~\bibnamefont {Li}}, \bibinfo {author}
  {\bibfnamefont {J.}~\bibnamefont {Fan}}, \bibinfo {author} {\bibfnamefont
  {X.}~\bibnamefont {Zhou}}, \bibinfo {author} {\bibfnamefont {X.}~\bibnamefont
  {Lv}}, \bibinfo {author} {\bibfnamefont {X.}~\bibnamefont {Yang}}, \bibinfo
  {author} {\bibfnamefont {Z.}~\bibnamefont {Cheng}},\ and\ \bibinfo {author}
  {\bibfnamefont {R.}~\bibnamefont {Wang}},\ }\bibfield  {title} {\bibinfo
  {title} {Concurrent high thermal conductivity and high carrier mobility in
  tetragonal tantalum nitride},\ }\href {https://doi.org/10.1063/5.0259103}
  {\bibfield  {journal} {\bibinfo  {journal} {Appl. Phys. Rev.}\ }\textbf
  {\bibinfo {volume} {12}},\ \bibinfo {pages} {021419} (\bibinfo {year}
  {2025})}\BibitemShut {NoStop}%
\bibitem [{\citenamefont {Wei}\ \emph {et~al.}(2024)\citenamefont {Wei},
  \citenamefont {Jin}, \citenamefont {Zhou}, \citenamefont {Yang},
  \citenamefont {Wang},\ and\ \citenamefont {Zhou}}]{wei2024tensile}%
  \BibitemOpen
  \bibfield  {author} {\bibinfo {author} {\bibfnamefont {L.}~\bibnamefont
  {Wei}}, \bibinfo {author} {\bibfnamefont {X.}~\bibnamefont {Jin}}, \bibinfo
  {author} {\bibfnamefont {Z.}~\bibnamefont {Zhou}}, \bibinfo {author}
  {\bibfnamefont {X.}~\bibnamefont {Yang}}, \bibinfo {author} {\bibfnamefont
  {G.}~\bibnamefont {Wang}},\ and\ \bibinfo {author} {\bibfnamefont
  {X.}~\bibnamefont {Zhou}},\ }\bibfield  {title} {\bibinfo {title} {Tensile
  strain induced enhancement of lattice thermal conductivity and its origin in
  two-dimensional {SnC}},\ }\href {https://doi.org/10.1103/PhysRevB.110.045406}
  {\bibfield  {journal} {\bibinfo  {journal} {Phys. Rev. B}\ }\textbf {\bibinfo
  {volume} {110}},\ \bibinfo {pages} {045406} (\bibinfo {year}
  {2024})}\BibitemShut {NoStop}%
\bibitem [{\citenamefont {Li}\ \emph {et~al.}(2018)\citenamefont {Li},
  \citenamefont {Ravichandran}, \citenamefont {Lindsay},\ and\ \citenamefont
  {Broido}}]{PhysRevLett.121.175901}%
  \BibitemOpen
  \bibfield  {author} {\bibinfo {author} {\bibfnamefont {C.}~\bibnamefont
  {Li}}, \bibinfo {author} {\bibfnamefont {N.~K.}\ \bibnamefont
  {Ravichandran}}, \bibinfo {author} {\bibfnamefont {L.}~\bibnamefont
  {Lindsay}},\ and\ \bibinfo {author} {\bibfnamefont {D.}~\bibnamefont
  {Broido}},\ }\bibfield  {title} {\bibinfo {title} {Fermi surface nesting and
  phonon frequency gap drive anomalous thermal transport},\ }\href
  {https://doi.org/10.1103/PhysRevLett.121.175901} {\bibfield  {journal}
  {\bibinfo  {journal} {Phys. Rev. Lett.}\ }\textbf {\bibinfo {volume} {121}},\
  \bibinfo {pages} {175901} (\bibinfo {year} {2018})}\BibitemShut {NoStop}%
\bibitem [{\citenamefont {Poncé}\ \emph {et~al.}(2016)\citenamefont {Poncé},
  \citenamefont {Margine}, \citenamefont {Verdi},\ and\ \citenamefont
  {Giustino}}]{ponce2016epw}%
  \BibitemOpen
  \bibfield  {author} {\bibinfo {author} {\bibfnamefont {S.}~\bibnamefont
  {Poncé}}, \bibinfo {author} {\bibfnamefont {E.}~\bibnamefont {Margine}},
  \bibinfo {author} {\bibfnamefont {C.}~\bibnamefont {Verdi}},\ and\ \bibinfo
  {author} {\bibfnamefont {F.}~\bibnamefont {Giustino}},\ }\bibfield  {title}
  {\bibinfo {title} {{EPW}: Electron–phonon coupling, transport and
  superconducting properties using maximally localized wannier functions},\
  }\href {https://doi.org/https://doi.org/10.1016/j.cpc.2016.07.028} {\bibfield
   {journal} {\bibinfo  {journal} {Comput. Phys. Commun.}\ }\textbf {\bibinfo
  {volume} {209}},\ \bibinfo {pages} {116} (\bibinfo {year}
  {2016})}\BibitemShut {NoStop}%
\bibitem [{\citenamefont {Zhou}\ \emph {et~al.}(2021)\citenamefont {Zhou},
  \citenamefont {Park}, \citenamefont {Lu}, \citenamefont {Maliyov},
  \citenamefont {Tong},\ and\ \citenamefont {Bernardi}}]{zhou2021perturbo}%
  \BibitemOpen
  \bibfield  {author} {\bibinfo {author} {\bibfnamefont {J.-J.}\ \bibnamefont
  {Zhou}}, \bibinfo {author} {\bibfnamefont {J.}~\bibnamefont {Park}}, \bibinfo
  {author} {\bibfnamefont {I.-T.}\ \bibnamefont {Lu}}, \bibinfo {author}
  {\bibfnamefont {I.}~\bibnamefont {Maliyov}}, \bibinfo {author} {\bibfnamefont
  {X.}~\bibnamefont {Tong}},\ and\ \bibinfo {author} {\bibfnamefont
  {M.}~\bibnamefont {Bernardi}},\ }\bibfield  {title} {\bibinfo {title}
  {Perturbo: A software package for ab initio electron–phonon interactions,
  charge transport and ultrafast dynamics},\ }\href
  {https://doi.org/https://doi.org/10.1016/j.cpc.2021.107970} {\bibfield
  {journal} {\bibinfo  {journal} {Comput. Phys. Commun.}\ }\textbf {\bibinfo
  {volume} {264}},\ \bibinfo {pages} {107970} (\bibinfo {year}
  {2021})}\BibitemShut {NoStop}%
\bibitem [{\citenamefont {Li}(2015)}]{li2015electrical}%
  \BibitemOpen
  \bibfield  {author} {\bibinfo {author} {\bibfnamefont {W.}~\bibnamefont
  {Li}},\ }\bibfield  {title} {\bibinfo {title} {Electrical transport limited
  by electron-phonon coupling from boltzmann transport equation: An ab initio
  study of {Si}, {Al}, and {${\mathrm{MoS}}_{2}$}},\ }\href
  {https://doi.org/10.1103/PhysRevB.92.075405} {\bibfield  {journal} {\bibinfo
  {journal} {Phys. Rev. B}\ }\textbf {\bibinfo {volume} {92}},\ \bibinfo
  {pages} {075405} (\bibinfo {year} {2015})}\BibitemShut {NoStop}%
\bibitem [{\citenamefont {Fiorentini}\ and\ \citenamefont
  {Bonini}(2016)}]{PhysRevB.94.085204}%
  \BibitemOpen
  \bibfield  {author} {\bibinfo {author} {\bibfnamefont {M.}~\bibnamefont
  {Fiorentini}}\ and\ \bibinfo {author} {\bibfnamefont {N.}~\bibnamefont
  {Bonini}},\ }\bibfield  {title} {\bibinfo {title} {Thermoelectric
  coefficients of $n$-doped silicon from first principles via the solution of
  the boltzmann transport equation},\ }\href
  {https://doi.org/10.1103/PhysRevB.94.085204} {\bibfield  {journal} {\bibinfo
  {journal} {Phys. Rev. B}\ }\textbf {\bibinfo {volume} {94}},\ \bibinfo
  {pages} {085204} (\bibinfo {year} {2016})}\BibitemShut {NoStop}%
\bibitem [{\citenamefont {Bl\"ochl}\ \emph {et~al.}(1994)\citenamefont
  {Bl\"ochl}, \citenamefont {Jepsen},\ and\ \citenamefont
  {Andersen}}]{PhysRevB.49.16223}%
  \BibitemOpen
  \bibfield  {author} {\bibinfo {author} {\bibfnamefont {P.~E.}\ \bibnamefont
  {Bl\"ochl}}, \bibinfo {author} {\bibfnamefont {O.}~\bibnamefont {Jepsen}},\
  and\ \bibinfo {author} {\bibfnamefont {O.~K.}\ \bibnamefont {Andersen}},\
  }\bibfield  {title} {\bibinfo {title} {Improved tetrahedron method for
  brillouin-zone integrations},\ }\href
  {https://doi.org/10.1103/PhysRevB.49.16223} {\bibfield  {journal} {\bibinfo
  {journal} {Phys. Rev. B}\ }\textbf {\bibinfo {volume} {49}},\ \bibinfo
  {pages} {16223} (\bibinfo {year} {1994})}\BibitemShut {NoStop}%
\bibitem [{\citenamefont {Kresse}\ and\ \citenamefont
  {Hafner}(1993)}]{kresse1993ab}%
  \BibitemOpen
  \bibfield  {author} {\bibinfo {author} {\bibfnamefont {G.}~\bibnamefont
  {Kresse}}\ and\ \bibinfo {author} {\bibfnamefont {J.}~\bibnamefont
  {Hafner}},\ }\bibfield  {title} {\bibinfo {title} {Ab initio molecular
  dynamics for open-shell transition metals},\ }\href
  {https://doi.org/10.1103/PhysRevB.48.13115} {\bibfield  {journal} {\bibinfo
  {journal} {Phys. Rev. B}\ }\textbf {\bibinfo {volume} {48}},\ \bibinfo
  {pages} {13115} (\bibinfo {year} {1993})}\BibitemShut {NoStop}%
\bibitem [{\citenamefont {Kresse}\ and\ \citenamefont
  {Furthm\"uller}(1996)}]{kresse1996efficient}%
  \BibitemOpen
  \bibfield  {author} {\bibinfo {author} {\bibfnamefont {G.}~\bibnamefont
  {Kresse}}\ and\ \bibinfo {author} {\bibfnamefont {J.}~\bibnamefont
  {Furthm\"uller}},\ }\bibfield  {title} {\bibinfo {title} {Efficient iterative
  schemes for ab initio total-energy calculations using a plane-wave basis
  set},\ }\href {https://doi.org/10.1103/PhysRevB.54.11169} {\bibfield
  {journal} {\bibinfo  {journal} {Phys. Rev. B}\ }\textbf {\bibinfo {volume}
  {54}},\ \bibinfo {pages} {11169} (\bibinfo {year} {1996})}\BibitemShut
  {NoStop}%
\bibitem [{\citenamefont {Bl\"ochl}(1994)}]{blochl1994projector}%
  \BibitemOpen
  \bibfield  {author} {\bibinfo {author} {\bibfnamefont {P.~E.}\ \bibnamefont
  {Bl\"ochl}},\ }\bibfield  {title} {\bibinfo {title} {Projector augmented-wave
  method},\ }\href {https://doi.org/10.1103/PhysRevB.50.17953} {\bibfield
  {journal} {\bibinfo  {journal} {Phys. Rev. B}\ }\textbf {\bibinfo {volume}
  {50}},\ \bibinfo {pages} {17953} (\bibinfo {year} {1994})}\BibitemShut
  {NoStop}%
\bibitem [{\citenamefont {Perdew}\ \emph {et~al.}(1996)\citenamefont {Perdew},
  \citenamefont {Burke},\ and\ \citenamefont
  {Ernzerhof}}]{perdew1996generalized}%
  \BibitemOpen
  \bibfield  {author} {\bibinfo {author} {\bibfnamefont {J.~P.}\ \bibnamefont
  {Perdew}}, \bibinfo {author} {\bibfnamefont {K.}~\bibnamefont {Burke}},\ and\
  \bibinfo {author} {\bibfnamefont {M.}~\bibnamefont {Ernzerhof}},\ }\bibfield
  {title} {\bibinfo {title} {Generalized gradient approximation made simple},\
  }\href {https://doi.org/10.1103/PhysRevLett.77.3865} {\bibfield  {journal}
  {\bibinfo  {journal} {Phys. Rev. Lett.}\ }\textbf {\bibinfo {volume} {77}},\
  \bibinfo {pages} {3865} (\bibinfo {year} {1996})}\BibitemShut {NoStop}%
\bibitem [{\citenamefont {Giannozzi}\ \emph {et~al.}(2009)\citenamefont
  {Giannozzi}, \citenamefont {Baroni}, \citenamefont {Bonini}, \citenamefont
  {Calandra}, \citenamefont {Car}, \citenamefont {Cavazzoni}, \citenamefont
  {Ceresoli}, \citenamefont {Chiarotti}, \citenamefont {Cococcioni},
  \citenamefont {Dabo}, \citenamefont {Dal~Corso}, \citenamefont
  {de~Gironcoli}, \citenamefont {Fabris}, \citenamefont {Fratesi},
  \citenamefont {Gebauer}, \citenamefont {Gerstmann}, \citenamefont
  {Gougoussis}, \citenamefont {Kokalj}, \citenamefont {Lazzeri}, \citenamefont
  {Martin-Samos}, \citenamefont {Marzari}, \citenamefont {Mauri}, \citenamefont
  {Mazzarello}, \citenamefont {Paolini}, \citenamefont {Pasquarello},
  \citenamefont {Paulatto}, \citenamefont {Sbraccia}, \citenamefont {Scandolo},
  \citenamefont {Sclauzero}, \citenamefont {Seitsonen}, \citenamefont
  {Smogunov}, \citenamefont {Umari},\ and\ \citenamefont
  {Wentzcovitch}}]{giannozzi2009quantum}%
  \BibitemOpen
  \bibfield  {author} {\bibinfo {author} {\bibfnamefont {P.}~\bibnamefont
  {Giannozzi}}, \bibinfo {author} {\bibfnamefont {S.}~\bibnamefont {Baroni}},
  \bibinfo {author} {\bibfnamefont {N.}~\bibnamefont {Bonini}}, \bibinfo
  {author} {\bibfnamefont {M.}~\bibnamefont {Calandra}}, \bibinfo {author}
  {\bibfnamefont {R.}~\bibnamefont {Car}}, \bibinfo {author} {\bibfnamefont
  {C.}~\bibnamefont {Cavazzoni}}, \bibinfo {author} {\bibfnamefont
  {D.}~\bibnamefont {Ceresoli}}, \bibinfo {author} {\bibfnamefont {G.~L.}\
  \bibnamefont {Chiarotti}}, \bibinfo {author} {\bibfnamefont {M.}~\bibnamefont
  {Cococcioni}}, \bibinfo {author} {\bibfnamefont {I.}~\bibnamefont {Dabo}},
  \bibinfo {author} {\bibfnamefont {A.}~\bibnamefont {Dal~Corso}}, \bibinfo
  {author} {\bibfnamefont {S.}~\bibnamefont {de~Gironcoli}}, \bibinfo {author}
  {\bibfnamefont {S.}~\bibnamefont {Fabris}}, \bibinfo {author} {\bibfnamefont
  {G.}~\bibnamefont {Fratesi}}, \bibinfo {author} {\bibfnamefont
  {R.}~\bibnamefont {Gebauer}}, \bibinfo {author} {\bibfnamefont
  {U.}~\bibnamefont {Gerstmann}}, \bibinfo {author} {\bibfnamefont
  {C.}~\bibnamefont {Gougoussis}}, \bibinfo {author} {\bibfnamefont
  {A.}~\bibnamefont {Kokalj}}, \bibinfo {author} {\bibfnamefont
  {M.}~\bibnamefont {Lazzeri}}, \bibinfo {author} {\bibfnamefont
  {L.}~\bibnamefont {Martin-Samos}}, \bibinfo {author} {\bibfnamefont
  {N.}~\bibnamefont {Marzari}}, \bibinfo {author} {\bibfnamefont
  {F.}~\bibnamefont {Mauri}}, \bibinfo {author} {\bibfnamefont
  {R.}~\bibnamefont {Mazzarello}}, \bibinfo {author} {\bibfnamefont
  {S.}~\bibnamefont {Paolini}}, \bibinfo {author} {\bibfnamefont
  {A.}~\bibnamefont {Pasquarello}}, \bibinfo {author} {\bibfnamefont
  {L.}~\bibnamefont {Paulatto}}, \bibinfo {author} {\bibfnamefont
  {C.}~\bibnamefont {Sbraccia}}, \bibinfo {author} {\bibfnamefont
  {S.}~\bibnamefont {Scandolo}}, \bibinfo {author} {\bibfnamefont
  {G.}~\bibnamefont {Sclauzero}}, \bibinfo {author} {\bibfnamefont {A.~P.}\
  \bibnamefont {Seitsonen}}, \bibinfo {author} {\bibfnamefont {A.}~\bibnamefont
  {Smogunov}}, \bibinfo {author} {\bibfnamefont {P.}~\bibnamefont {Umari}},\
  and\ \bibinfo {author} {\bibfnamefont {R.~M.}\ \bibnamefont {Wentzcovitch}},\
  }\bibfield  {title} {\bibinfo {title} {Quantum espresso: a modular and
  open-source software project for quantum simulations of materials},\ }\href
  {https://doi.org/10.1088/0953-8984/21/39/395502} {\bibfield  {journal}
  {\bibinfo  {journal} {J. Phys. Condens. Matter}\ }\textbf {\bibinfo {volume}
  {21}},\ \bibinfo {pages} {395502} (\bibinfo {year} {2009})}\BibitemShut
  {NoStop}%
\bibitem [{\citenamefont {Li}\ \emph {et~al.}(2014)\citenamefont {Li},
  \citenamefont {Carrete}, \citenamefont {{A. Katcho}},\ and\ \citenamefont
  {Mingo}}]{li2014shengbte}%
  \BibitemOpen
  \bibfield  {author} {\bibinfo {author} {\bibfnamefont {W.}~\bibnamefont
  {Li}}, \bibinfo {author} {\bibfnamefont {J.}~\bibnamefont {Carrete}},
  \bibinfo {author} {\bibfnamefont {N.}~\bibnamefont {{A. Katcho}}},\ and\
  \bibinfo {author} {\bibfnamefont {N.}~\bibnamefont {Mingo}},\ }\bibfield
  {title} {\bibinfo {title} {Shengbte: A solver of the boltzmann transport
  equation for phonons},\ }\href
  {https://doi.org/https://doi.org/10.1016/j.cpc.2014.02.015} {\bibfield
  {journal} {\bibinfo  {journal} {Comput. Phys. Commun.}\ }\textbf {\bibinfo
  {volume} {185}},\ \bibinfo {pages} {1747} (\bibinfo {year}
  {2014})}\BibitemShut {NoStop}%
\bibitem [{\citenamefont {Han}\ \emph {et~al.}(2022)\citenamefont {Han},
  \citenamefont {Yang}, \citenamefont {Li}, \citenamefont {Feng},\ and\
  \citenamefont {Ruan}}]{han2022fourphonon}%
  \BibitemOpen
  \bibfield  {author} {\bibinfo {author} {\bibfnamefont {Z.}~\bibnamefont
  {Han}}, \bibinfo {author} {\bibfnamefont {X.}~\bibnamefont {Yang}}, \bibinfo
  {author} {\bibfnamefont {W.}~\bibnamefont {Li}}, \bibinfo {author}
  {\bibfnamefont {T.}~\bibnamefont {Feng}},\ and\ \bibinfo {author}
  {\bibfnamefont {X.}~\bibnamefont {Ruan}},\ }\bibfield  {title} {\bibinfo
  {title} {{FourPhonon}: An extension module to shengbte for computing
  four-phonon scattering rates and thermal conductivity},\ }\href
  {https://doi.org/https://doi.org/10.1016/j.cpc.2021.108179} {\bibfield
  {journal} {\bibinfo  {journal} {Comput. Phys. Commun.}\ }\textbf {\bibinfo
  {volume} {270}},\ \bibinfo {pages} {108179} (\bibinfo {year}
  {2022})}\BibitemShut {NoStop}%
\bibitem [{\citenamefont {Yang}\ \emph
  {et~al.}(2021{\natexlab{a}})\citenamefont {Yang}, \citenamefont {Jena},
  \citenamefont {Meng}, \citenamefont {Wen}, \citenamefont {Ma}, \citenamefont
  {Li},\ and\ \citenamefont {Li}}]{yang2021indirect}%
  \BibitemOpen
  \bibfield  {author} {\bibinfo {author} {\bibfnamefont {X.}~\bibnamefont
  {Yang}}, \bibinfo {author} {\bibfnamefont {A.}~\bibnamefont {Jena}}, \bibinfo
  {author} {\bibfnamefont {F.}~\bibnamefont {Meng}}, \bibinfo {author}
  {\bibfnamefont {S.}~\bibnamefont {Wen}}, \bibinfo {author} {\bibfnamefont
  {J.}~\bibnamefont {Ma}}, \bibinfo {author} {\bibfnamefont {X.}~\bibnamefont
  {Li}},\ and\ \bibinfo {author} {\bibfnamefont {W.}~\bibnamefont {Li}},\
  }\bibfield  {title} {\bibinfo {title} {Indirect electron-phonon interaction
  leading to significant reduction of thermal conductivity in graphene},\
  }\href {https://doi.org/https://doi.org/10.1016/j.mtphys.2020.100315}
  {\bibfield  {journal} {\bibinfo  {journal} {Mat. Today Phys.}\ }\textbf
  {\bibinfo {volume} {18}},\ \bibinfo {pages} {100315} (\bibinfo {year}
  {2021}{\natexlab{a}})}\BibitemShut {NoStop}%
\bibitem [{\citenamefont {Yang}\ \emph
  {et~al.}(2021{\natexlab{b}})\citenamefont {Yang}, \citenamefont {Liu},
  \citenamefont {Meng},\ and\ \citenamefont {Li}}]{yang2021tuning}%
  \BibitemOpen
  \bibfield  {author} {\bibinfo {author} {\bibfnamefont {X.}~\bibnamefont
  {Yang}}, \bibinfo {author} {\bibfnamefont {Z.}~\bibnamefont {Liu}}, \bibinfo
  {author} {\bibfnamefont {F.}~\bibnamefont {Meng}},\ and\ \bibinfo {author}
  {\bibfnamefont {W.}~\bibnamefont {Li}},\ }\bibfield  {title} {\bibinfo
  {title} {Tuning the phonon transport in bilayer graphene to an anomalous
  regime dominated by electron-phonon scattering},\ }\href
  {https://doi.org/10.1103/PhysRevB.104.L100306} {\bibfield  {journal}
  {\bibinfo  {journal} {Phys. Rev. B}\ }\textbf {\bibinfo {volume} {104}},\
  \bibinfo {pages} {L100306} (\bibinfo {year}
  {2021}{\natexlab{b}})}\BibitemShut {NoStop}%
\bibitem [{SM()}]{SM}%
  \BibitemOpen
  \href@noop {} {\bibinfo  {journal} {See supplemental material at [URL] for
  additional information about the computational details, convergence tests,
  and supplementary figures, which includes
  Refs.~\cite{kresse1993ab,kresse1996efficient,blochl1994projector,perdew1996generalized,li2014shengbte,giannozzi2009quantum,ponce2016epw,zhou2021perturbo}}\
  }\BibitemShut {NoStop}%
\bibitem [{\citenamefont {Lv}\ \emph {et~al.}(2015{\natexlab{b}})\citenamefont
  {Lv}, \citenamefont {Xu}, \citenamefont {Weng}, \citenamefont {Ma},
  \citenamefont {Richard}, \citenamefont {Huang}, \citenamefont {Zhao},
  \citenamefont {Chen}, \citenamefont {Matt}, \citenamefont {Bisti} \emph
  {et~al.}}]{lv2015observation}%
  \BibitemOpen
\bibfield  {journal} {  }\bibfield  {author} {\bibinfo {author} {\bibfnamefont
  {B.}~\bibnamefont {Lv}}, \bibinfo {author} {\bibfnamefont {N.}~\bibnamefont
  {Xu}}, \bibinfo {author} {\bibfnamefont {H.}~\bibnamefont {Weng}}, \bibinfo
  {author} {\bibfnamefont {J.}~\bibnamefont {Ma}}, \bibinfo {author}
  {\bibfnamefont {P.}~\bibnamefont {Richard}}, \bibinfo {author} {\bibfnamefont
  {X.}~\bibnamefont {Huang}}, \bibinfo {author} {\bibfnamefont
  {L.}~\bibnamefont {Zhao}}, \bibinfo {author} {\bibfnamefont {G.}~\bibnamefont
  {Chen}}, \bibinfo {author} {\bibfnamefont {C.}~\bibnamefont {Matt}}, \bibinfo
  {author} {\bibfnamefont {F.}~\bibnamefont {Bisti}}, \emph {et~al.},\
  }\bibfield  {title} {\bibinfo {title} {Observation of weyl nodes in {TaAs}},\
  }\href {https://doi.org/10.1038/nphys3426} {\bibfield  {journal} {\bibinfo
  {journal} {Nat. Phys.}\ }\textbf {\bibinfo {volume} {11}},\ \bibinfo {pages}
  {724} (\bibinfo {year} {2015}{\natexlab{b}})}\BibitemShut {NoStop}%
\bibitem [{\citenamefont {Liu}\ \emph {et~al.}(2025)\citenamefont {Liu},
  \citenamefont {Mishra}, \citenamefont {Lihm}, \citenamefont {Ponc\'e},\ and\
  \citenamefont {Margine}}]{x8zl-w5x3}%
  \BibitemOpen
  \bibfield  {author} {\bibinfo {author} {\bibfnamefont {Z.}~\bibnamefont
  {Liu}}, \bibinfo {author} {\bibfnamefont {S.~B.}\ \bibnamefont {Mishra}},
  \bibinfo {author} {\bibfnamefont {J.-M.}\ \bibnamefont {Lihm}}, \bibinfo
  {author} {\bibfnamefont {S.}~\bibnamefont {Ponc\'e}},\ and\ \bibinfo {author}
  {\bibfnamefont {E.~R.}\ \bibnamefont {Margine}},\ }\bibfield  {title}
  {\bibinfo {title} {Phonon-limited carrier transport in the weyl semimetal
  {TaAs}},\ }\href {https://doi.org/10.1103/x8zl-w5x3} {\bibfield  {journal}
  {\bibinfo  {journal} {Phys. Rev. B}\ }\textbf {\bibinfo {volume} {112}},\
  \bibinfo {pages} {104311} (\bibinfo {year} {2025})}\BibitemShut {NoStop}%
\bibitem [{\citenamefont {Lindsay}\ \emph {et~al.}(2013)\citenamefont
  {Lindsay}, \citenamefont {Broido},\ and\ \citenamefont
  {Reinecke}}]{lindsay2013first}%
  \BibitemOpen
  \bibfield  {author} {\bibinfo {author} {\bibfnamefont {L.}~\bibnamefont
  {Lindsay}}, \bibinfo {author} {\bibfnamefont {D.~A.}\ \bibnamefont
  {Broido}},\ and\ \bibinfo {author} {\bibfnamefont {T.~L.}\ \bibnamefont
  {Reinecke}},\ }\bibfield  {title} {\bibinfo {title} {First-principles
  determination of ultrahigh thermal conductivity of boron arsenide: A
  competitor for diamond?},\ }\href
  {https://doi.org/10.1103/PhysRevLett.111.025901} {\bibfield  {journal}
  {\bibinfo  {journal} {Phys. Rev. Lett.}\ }\textbf {\bibinfo {volume} {111}},\
  \bibinfo {pages} {025901} (\bibinfo {year} {2013})}\BibitemShut {NoStop}%
\bibitem [{\citenamefont {Lee}\ \emph {et~al.}(2015)\citenamefont {Lee},
  \citenamefont {Xu}, \citenamefont {Huang}, \citenamefont {Sanchez},
  \citenamefont {Belopolski}, \citenamefont {Chang}, \citenamefont {Bian},
  \citenamefont {Alidoust}, \citenamefont {Zheng}, \citenamefont {Neupane},
  \citenamefont {Wang}, \citenamefont {Bansil}, \citenamefont {Hasan},\ and\
  \citenamefont {Lin}}]{lee2015fermi}%
  \BibitemOpen
  \bibfield  {author} {\bibinfo {author} {\bibfnamefont {C.-C.}\ \bibnamefont
  {Lee}}, \bibinfo {author} {\bibfnamefont {S.-Y.}\ \bibnamefont {Xu}},
  \bibinfo {author} {\bibfnamefont {S.-M.}\ \bibnamefont {Huang}}, \bibinfo
  {author} {\bibfnamefont {D.~S.}\ \bibnamefont {Sanchez}}, \bibinfo {author}
  {\bibfnamefont {I.}~\bibnamefont {Belopolski}}, \bibinfo {author}
  {\bibfnamefont {G.}~\bibnamefont {Chang}}, \bibinfo {author} {\bibfnamefont
  {G.}~\bibnamefont {Bian}}, \bibinfo {author} {\bibfnamefont {N.}~\bibnamefont
  {Alidoust}}, \bibinfo {author} {\bibfnamefont {H.}~\bibnamefont {Zheng}},
  \bibinfo {author} {\bibfnamefont {M.}~\bibnamefont {Neupane}}, \bibinfo
  {author} {\bibfnamefont {B.}~\bibnamefont {Wang}}, \bibinfo {author}
  {\bibfnamefont {A.}~\bibnamefont {Bansil}}, \bibinfo {author} {\bibfnamefont
  {M.~Z.}\ \bibnamefont {Hasan}},\ and\ \bibinfo {author} {\bibfnamefont
  {H.}~\bibnamefont {Lin}},\ }\bibfield  {title} {\bibinfo {title} {Fermi
  surface interconnectivity and topology in weyl fermion semimetals {TaAs},
  {TaP}, {NbAs}, and {NbP}},\ }\href
  {https://doi.org/10.1103/PhysRevB.92.235104} {\bibfield  {journal} {\bibinfo
  {journal} {Phys. Rev. B}\ }\textbf {\bibinfo {volume} {92}},\ \bibinfo
  {pages} {235104} (\bibinfo {year} {2015})}\BibitemShut {NoStop}%
\bibitem [{\citenamefont {Peng}\ \emph {et~al.}(2016)\citenamefont {Peng},
  \citenamefont {Zhang}, \citenamefont {Shao}, \citenamefont {Lu},
  \citenamefont {Zhang},\ and\ \citenamefont {Zhu}}]{PENG2016225}%
  \BibitemOpen
  \bibfield  {author} {\bibinfo {author} {\bibfnamefont {B.}~\bibnamefont
  {Peng}}, \bibinfo {author} {\bibfnamefont {H.}~\bibnamefont {Zhang}},
  \bibinfo {author} {\bibfnamefont {H.}~\bibnamefont {Shao}}, \bibinfo {author}
  {\bibfnamefont {H.}~\bibnamefont {Lu}}, \bibinfo {author} {\bibfnamefont
  {D.~W.}\ \bibnamefont {Zhang}},\ and\ \bibinfo {author} {\bibfnamefont
  {H.}~\bibnamefont {Zhu}},\ }\bibfield  {title} {\bibinfo {title} {High
  thermoelectric performance of weyl semimetal {TaAs}},\ }\href
  {https://doi.org/https://doi.org/10.1016/j.nanoen.2016.10.016} {\bibfield
  {journal} {\bibinfo  {journal} {Nano Energy}\ }\textbf {\bibinfo {volume}
  {30}},\ \bibinfo {pages} {225} (\bibinfo {year} {2016})}\BibitemShut
  {NoStop}%
\bibitem [{\citenamefont {Ouyang}\ \emph {et~al.}(2016)\citenamefont {Ouyang},
  \citenamefont {Xiao}, \citenamefont {Tang}, \citenamefont {Hu},\ and\
  \citenamefont {Zhong}}]{C6CP02935C}%
  \BibitemOpen
  \bibfield  {author} {\bibinfo {author} {\bibfnamefont {T.}~\bibnamefont
  {Ouyang}}, \bibinfo {author} {\bibfnamefont {H.}~\bibnamefont {Xiao}},
  \bibinfo {author} {\bibfnamefont {C.}~\bibnamefont {Tang}}, \bibinfo {author}
  {\bibfnamefont {M.}~\bibnamefont {Hu}},\ and\ \bibinfo {author}
  {\bibfnamefont {J.}~\bibnamefont {Zhong}},\ }\bibfield  {title} {\bibinfo
  {title} {Anisotropic thermal transport in weyl semimetal {TaAs}: a first
  principles calculation},\ }\href {https://doi.org/10.1039/C6CP02935C}
  {\bibfield  {journal} {\bibinfo  {journal} {Phys. Chem. Chem. Phys.}\
  }\textbf {\bibinfo {volume} {18}},\ \bibinfo {pages} {16709} (\bibinfo {year}
  {2016})}\BibitemShut {NoStop}%
\bibitem [{\citenamefont {Xiang}\ \emph {et~al.}(2017)\citenamefont {Xiang},
  \citenamefont {Hu}, \citenamefont {Lv}, \citenamefont {Zhang}, \citenamefont
  {Zhao}, \citenamefont {Chen}, \citenamefont {Li}, \citenamefont {Chen},\ and\
  \citenamefont {Sun}}]{xiang2017anisotropic}%
  \BibitemOpen
  \bibfield  {author} {\bibinfo {author} {\bibfnamefont {J.}~\bibnamefont
  {Xiang}}, \bibinfo {author} {\bibfnamefont {S.}~\bibnamefont {Hu}}, \bibinfo
  {author} {\bibfnamefont {M.}~\bibnamefont {Lv}}, \bibinfo {author}
  {\bibfnamefont {J.}~\bibnamefont {Zhang}}, \bibinfo {author} {\bibfnamefont
  {H.}~\bibnamefont {Zhao}}, \bibinfo {author} {\bibfnamefont {G.}~\bibnamefont
  {Chen}}, \bibinfo {author} {\bibfnamefont {W.}~\bibnamefont {Li}}, \bibinfo
  {author} {\bibfnamefont {Z.}~\bibnamefont {Chen}},\ and\ \bibinfo {author}
  {\bibfnamefont {P.}~\bibnamefont {Sun}},\ }\bibfield  {title} {\bibinfo
  {title} {Anisotropic thermal and electrical transport of weyl semimetal
  {TaAs}},\ }\href {https://doi.org/10.1088/1361-648X/aa964b} {\bibfield
  {journal} {\bibinfo  {journal} {J. Phys. Condens. Matter}\ }\textbf {\bibinfo
  {volume} {29}},\ \bibinfo {pages} {485501} (\bibinfo {year}
  {2017})}\BibitemShut {NoStop}%
\bibitem [{\citenamefont {Han}\ \emph {et~al.}(2023)\citenamefont {Han},
  \citenamefont {Tang}, \citenamefont {Yuan}, \citenamefont {Luo},\ and\
  \citenamefont {Liu}}]{HAN2023520}%
  \BibitemOpen
  \bibfield  {author} {\bibinfo {author} {\bibfnamefont {S.}~\bibnamefont
  {Han}}, \bibinfo {author} {\bibfnamefont {Q.}~\bibnamefont {Tang}}, \bibinfo
  {author} {\bibfnamefont {H.}~\bibnamefont {Yuan}}, \bibinfo {author}
  {\bibfnamefont {Y.}~\bibnamefont {Luo}},\ and\ \bibinfo {author}
  {\bibfnamefont {H.}~\bibnamefont {Liu}},\ }\bibfield  {title} {\bibinfo
  {title} {Effects of electron-phonon coupling on the phonon transport
  properties of the weyl semimetals {NbAs} and {TaAs}: A comparative study},\
  }\href {https://doi.org/https://doi.org/10.1016/j.jmat.2022.12.001}
  {\bibfield  {journal} {\bibinfo  {journal} {J. Materiomics}\ }\textbf
  {\bibinfo {volume} {9}},\ \bibinfo {pages} {520} (\bibinfo {year}
  {2023})}\BibitemShut {NoStop}%
\bibitem [{\citenamefont {Chen}\ \emph {et~al.}(2025)\citenamefont {Chen},
  \citenamefont {Jin},\ and\ \citenamefont {Yang}}]{chen2025symmetry}%
  \BibitemOpen
  \bibfield  {author} {\bibinfo {author} {\bibfnamefont {K.}~\bibnamefont
  {Chen}}, \bibinfo {author} {\bibfnamefont {X.}~\bibnamefont {Jin}},\ and\
  \bibinfo {author} {\bibfnamefont {X.}~\bibnamefont {Yang}},\ }\bibfield
  {title} {\bibinfo {title} {Symmetry-breaking strain drives significant
  reduction in lattice thermal conductivity: A case study of boron arsenide},\
  }\href {https://doi.org/10.1088/0256-307X/42/12/120801} {\bibfield  {journal}
  {\bibinfo  {journal} {Chin. Phys. Lett.}\ }\textbf {\bibinfo {volume} {42}},\
  \bibinfo {pages} {120801} (\bibinfo {year} {2025})}\BibitemShut {NoStop}%
\bibitem [{\citenamefont {Wang}\ \emph {et~al.}(2025)\citenamefont {Wang},
  \citenamefont {Li}, \citenamefont {Ju}, \citenamefont {Zhang}, \citenamefont
  {Ma}, \citenamefont {Li},\ and\ \citenamefont {Zhang}}]{wang2025atomic}%
  \BibitemOpen
  \bibfield  {author} {\bibinfo {author} {\bibfnamefont {T.}~\bibnamefont
  {Wang}}, \bibinfo {author} {\bibfnamefont {X.}~\bibnamefont {Li}}, \bibinfo
  {author} {\bibfnamefont {Z.}~\bibnamefont {Ju}}, \bibinfo {author}
  {\bibfnamefont {G.}~\bibnamefont {Zhang}}, \bibinfo {author} {\bibfnamefont
  {D.}~\bibnamefont {Ma}}, \bibinfo {author} {\bibfnamefont {W.}~\bibnamefont
  {Li}},\ and\ \bibinfo {author} {\bibfnamefont {L.}~\bibnamefont {Zhang}},\
  }\bibfield  {title} {\bibinfo {title} {Atomic mass engineering of ultra-high
  thermal conductivity in large bandgap materials: A case study with boron
  arsenide},\ }\href {https://doi.org/10.1088/0256-307X/42/7/070802} {\bibfield
   {journal} {\bibinfo  {journal} {Chin. Phys. Lett.}\ }\textbf {\bibinfo
  {volume} {42}},\ \bibinfo {pages} {070802} (\bibinfo {year}
  {2025})}\BibitemShut {NoStop}%
\bibitem [{\citenamefont {Witczak-Krempa}\ and\ \citenamefont
  {Kim}(2012)}]{witczak2012topological}%
  \BibitemOpen
  \bibfield  {author} {\bibinfo {author} {\bibfnamefont {W.}~\bibnamefont
  {Witczak-Krempa}}\ and\ \bibinfo {author} {\bibfnamefont {Y.~B.}\
  \bibnamefont {Kim}},\ }\bibfield  {title} {\bibinfo {title} {Topological and
  magnetic phases of interacting electrons in the pyrochlore iridates},\ }\href
  {https://doi.org/10.1103/PhysRevB.85.045124} {\bibfield  {journal} {\bibinfo
  {journal} {Phys. Rev. B}\ }\textbf {\bibinfo {volume} {85}},\ \bibinfo
  {pages} {045124} (\bibinfo {year} {2012})}\BibitemShut {NoStop}%
\bibitem [{\citenamefont {Zhang}\ \emph {et~al.}(2017)\citenamefont {Zhang},
  \citenamefont {Yuan}, \citenamefont {Jiang}, \citenamefont {Tong},
  \citenamefont {Zhang}, \citenamefont {Xie},\ and\ \citenamefont
  {Jia}}]{zhang2017electron}%
  \BibitemOpen
  \bibfield  {author} {\bibinfo {author} {\bibfnamefont {C.-L.}\ \bibnamefont
  {Zhang}}, \bibinfo {author} {\bibfnamefont {Z.}~\bibnamefont {Yuan}},
  \bibinfo {author} {\bibfnamefont {Q.-D.}\ \bibnamefont {Jiang}}, \bibinfo
  {author} {\bibfnamefont {B.}~\bibnamefont {Tong}}, \bibinfo {author}
  {\bibfnamefont {C.}~\bibnamefont {Zhang}}, \bibinfo {author} {\bibfnamefont
  {X.~C.}\ \bibnamefont {Xie}},\ and\ \bibinfo {author} {\bibfnamefont
  {S.}~\bibnamefont {Jia}},\ }\bibfield  {title} {\bibinfo {title} {Electron
  scattering in tantalum monoarsenide},\ }\href
  {https://doi.org/10.1103/PhysRevB.95.085202} {\bibfield  {journal} {\bibinfo
  {journal} {Phys. Rev. B}\ }\textbf {\bibinfo {volume} {95}},\ \bibinfo
  {pages} {085202} (\bibinfo {year} {2017})}\BibitemShut {NoStop}%
\bibitem [{\citenamefont {Allemand}\ \emph {et~al.}(2025)\citenamefont
  {Allemand}, \citenamefont {Giantomassi},\ and\ \citenamefont
  {Verstraete}}]{7lnq-snmp}%
  \BibitemOpen
  \bibfield  {author} {\bibinfo {author} {\bibfnamefont {G.~E.}\ \bibnamefont
  {Allemand}}, \bibinfo {author} {\bibfnamefont {M.}~\bibnamefont
  {Giantomassi}},\ and\ \bibinfo {author} {\bibfnamefont {M.~J.}\ \bibnamefont
  {Verstraete}},\ }\bibfield  {title} {\bibinfo {title} {First-principles
  calculations of transport coefficients in the weyl semimetal {TaAs}},\ }\href
  {https://doi.org/10.1103/7lnq-snmp} {\bibfield  {journal} {\bibinfo
  {journal} {Phys. Rev. B}\ }\textbf {\bibinfo {volume} {112}},\ \bibinfo
  {pages} {125122} (\bibinfo {year} {2025})}\BibitemShut {NoStop}%
\bibitem [{\citenamefont {Zhang}\ \emph {et~al.}(2015)\citenamefont {Zhang},
  \citenamefont {Guo}, \citenamefont {Lu}, \citenamefont {Zhang}, \citenamefont
  {Yuan}, \citenamefont {Lin}, \citenamefont {Wang},\ and\ \citenamefont
  {Jia}}]{PhysRevB.92.041203}%
  \BibitemOpen
  \bibfield  {author} {\bibinfo {author} {\bibfnamefont {C.}~\bibnamefont
  {Zhang}}, \bibinfo {author} {\bibfnamefont {C.}~\bibnamefont {Guo}}, \bibinfo
  {author} {\bibfnamefont {H.}~\bibnamefont {Lu}}, \bibinfo {author}
  {\bibfnamefont {X.}~\bibnamefont {Zhang}}, \bibinfo {author} {\bibfnamefont
  {Z.}~\bibnamefont {Yuan}}, \bibinfo {author} {\bibfnamefont {Z.}~\bibnamefont
  {Lin}}, \bibinfo {author} {\bibfnamefont {J.}~\bibnamefont {Wang}},\ and\
  \bibinfo {author} {\bibfnamefont {S.}~\bibnamefont {Jia}},\ }\bibfield
  {title} {\bibinfo {title} {Large magnetoresistance over an extended
  temperature regime in monophosphides of tantalum and niobium},\ }\href
  {https://doi.org/10.1103/PhysRevB.92.041203} {\bibfield  {journal} {\bibinfo
  {journal} {Phys. Rev. B}\ }\textbf {\bibinfo {volume} {92}},\ \bibinfo
  {pages} {041203(R)} (\bibinfo {year} {2015})}\BibitemShut {NoStop}%
\bibitem [{\citenamefont {Mishra}\ \emph {et~al.}(2026)\citenamefont {Mishra},
  \citenamefont {Liu}, \citenamefont {Tiwari}, \citenamefont {Giustino},\ and\
  \citenamefont {Margine}}]{d9np-11j1}%
  \BibitemOpen
  \bibfield  {author} {\bibinfo {author} {\bibfnamefont {S.~B.}\ \bibnamefont
  {Mishra}}, \bibinfo {author} {\bibfnamefont {Z.}~\bibnamefont {Liu}},
  \bibinfo {author} {\bibfnamefont {S.}~\bibnamefont {Tiwari}}, \bibinfo
  {author} {\bibfnamefont {F.}~\bibnamefont {Giustino}},\ and\ \bibinfo
  {author} {\bibfnamefont {E.~R.}\ \bibnamefont {Margine}},\ }\bibfield
  {title} {\bibinfo {title} {Comparative study of phonon-limited carrier
  transport in the weyl semimetal {TaAs} family},\ }\href
  {https://doi.org/10.1103/d9np-11j1} {\bibfield  {journal} {\bibinfo
  {journal} {Phys. Rev. B}\ }\textbf {\bibinfo {volume} {113}},\ \bibinfo
  {pages} {045202} (\bibinfo {year} {2026})}\BibitemShut {NoStop}%
\bibitem [{\citenamefont {Li}\ \emph {et~al.}(2020)\citenamefont {Li},
  \citenamefont {Tong}, \citenamefont {Zhang},\ and\ \citenamefont
  {Bao}}]{PhysRevB.102.174306}%
  \BibitemOpen
  \bibfield  {author} {\bibinfo {author} {\bibfnamefont {S.}~\bibnamefont
  {Li}}, \bibinfo {author} {\bibfnamefont {Z.}~\bibnamefont {Tong}}, \bibinfo
  {author} {\bibfnamefont {X.}~\bibnamefont {Zhang}},\ and\ \bibinfo {author}
  {\bibfnamefont {H.}~\bibnamefont {Bao}},\ }\bibfield  {title} {\bibinfo
  {title} {Thermal conductivity and lorenz ratio of metals at intermediate
  temperatures with mode-level first-principles analysis},\ }\href
  {https://doi.org/10.1103/PhysRevB.102.174306} {\bibfield  {journal} {\bibinfo
   {journal} {Phys. Rev. B}\ }\textbf {\bibinfo {volume} {102}},\ \bibinfo
  {pages} {174306} (\bibinfo {year} {2020})}\BibitemShut {NoStop}%
\bibitem [{\citenamefont {Li}\ \emph {et~al.}(2019)\citenamefont {Li},
  \citenamefont {Li}, \citenamefont {Wang}, \citenamefont {Liu}, \citenamefont
  {Ma}, \citenamefont {Song}, \citenamefont {Li}, \citenamefont {Li},\ and\
  \citenamefont {Chen}}]{PhysRevLett.123.136802}%
  \BibitemOpen
  \bibfield  {author} {\bibinfo {author} {\bibfnamefont {R.}~\bibnamefont
  {Li}}, \bibinfo {author} {\bibfnamefont {J.}~\bibnamefont {Li}}, \bibinfo
  {author} {\bibfnamefont {L.}~\bibnamefont {Wang}}, \bibinfo {author}
  {\bibfnamefont {J.}~\bibnamefont {Liu}}, \bibinfo {author} {\bibfnamefont
  {H.}~\bibnamefont {Ma}}, \bibinfo {author} {\bibfnamefont {H.-F.}\
  \bibnamefont {Song}}, \bibinfo {author} {\bibfnamefont {D.}~\bibnamefont
  {Li}}, \bibinfo {author} {\bibfnamefont {Y.}~\bibnamefont {Li}},\ and\
  \bibinfo {author} {\bibfnamefont {X.-Q.}\ \bibnamefont {Chen}},\ }\bibfield
  {title} {\bibinfo {title} {Underlying topological dirac nodal line mechanism
  of the anomalously large electron-phonon coupling strength on a be (0001)
  surface},\ }\href {https://doi.org/10.1103/PhysRevLett.123.136802} {\bibfield
   {journal} {\bibinfo  {journal} {Phys. Rev. Lett.}\ }\textbf {\bibinfo
  {volume} {123}},\ \bibinfo {pages} {136802} (\bibinfo {year}
  {2019})}\BibitemShut {NoStop}%
\bibitem [{\citenamefont {Jain}\ \emph {et~al.}(2013)\citenamefont {Jain},
  \citenamefont {Ong}, \citenamefont {Hautier}, \citenamefont {Chen},
  \citenamefont {Richards}, \citenamefont {Dacek}, \citenamefont {Cholia},
  \citenamefont {Gunter}, \citenamefont {Skinner}, \citenamefont {Ceder},\ and\
  \citenamefont {Persson}}]{10.1063/1.4812323}%
  \BibitemOpen
  \bibfield  {author} {\bibinfo {author} {\bibfnamefont {A.}~\bibnamefont
  {Jain}}, \bibinfo {author} {\bibfnamefont {S.~P.}\ \bibnamefont {Ong}},
  \bibinfo {author} {\bibfnamefont {G.}~\bibnamefont {Hautier}}, \bibinfo
  {author} {\bibfnamefont {W.}~\bibnamefont {Chen}}, \bibinfo {author}
  {\bibfnamefont {W.~D.}\ \bibnamefont {Richards}}, \bibinfo {author}
  {\bibfnamefont {S.}~\bibnamefont {Dacek}}, \bibinfo {author} {\bibfnamefont
  {S.}~\bibnamefont {Cholia}}, \bibinfo {author} {\bibfnamefont
  {D.}~\bibnamefont {Gunter}}, \bibinfo {author} {\bibfnamefont
  {D.}~\bibnamefont {Skinner}}, \bibinfo {author} {\bibfnamefont
  {G.}~\bibnamefont {Ceder}},\ and\ \bibinfo {author} {\bibfnamefont {K.~A.}\
  \bibnamefont {Persson}},\ }\bibfield  {title} {\bibinfo {title} {Commentary:
  The materials project: A materials genome approach to accelerating materials
  innovation},\ }\href {https://doi.org/10.1063/1.4812323} {\bibfield
  {journal} {\bibinfo  {journal} {APL Mater.}\ }\textbf {\bibinfo {volume}
  {1}},\ \bibinfo {pages} {011002} (\bibinfo {year} {2013})}\BibitemShut
  {NoStop}%
\bibitem [{\citenamefont {Munro}\ \emph {et~al.}(2020)\citenamefont {Munro},
  \citenamefont {Latimer}, \citenamefont {Horton}, \citenamefont {Dwaraknath},\
  and\ \citenamefont {Persson}}]{munro2020improved}%
  \BibitemOpen
  \bibfield  {author} {\bibinfo {author} {\bibfnamefont {J.~M.}\ \bibnamefont
  {Munro}}, \bibinfo {author} {\bibfnamefont {K.}~\bibnamefont {Latimer}},
  \bibinfo {author} {\bibfnamefont {M.~K.}\ \bibnamefont {Horton}}, \bibinfo
  {author} {\bibfnamefont {S.}~\bibnamefont {Dwaraknath}},\ and\ \bibinfo
  {author} {\bibfnamefont {K.~A.}\ \bibnamefont {Persson}},\ }\bibfield
  {title} {\bibinfo {title} {An improved symmetry-based approach to reciprocal
  space path selection in band structure calculations},\ }\href
  {https://doi.org/10.1038/s41524-020-00383-7} {\bibfield  {journal} {\bibinfo
  {journal} {npj Comput. Mater.}\ }\textbf {\bibinfo {volume} {6}},\ \bibinfo
  {pages} {112} (\bibinfo {year} {2020})}\BibitemShut {NoStop}%
\bibitem [{\citenamefont {Zhong}\ \emph {et~al.}(2026)\citenamefont {Zhong},
  \citenamefont {Jin}, \citenamefont {He}, \citenamefont {Wang}, \citenamefont
  {Zhou}, \citenamefont {Deng},\ and\ \citenamefont {Yang}}]{5svq-g1k7}%
  \BibitemOpen
  \bibfield  {author} {\bibinfo {author} {\bibfnamefont {L.}~\bibnamefont
  {Zhong}}, \bibinfo {author} {\bibfnamefont {X.}~\bibnamefont {Jin}}, \bibinfo
  {author} {\bibfnamefont {M.}~\bibnamefont {He}}, \bibinfo {author}
  {\bibfnamefont {R.}~\bibnamefont {Wang}}, \bibinfo {author} {\bibfnamefont
  {X.}~\bibnamefont {Zhou}}, \bibinfo {author} {\bibfnamefont {T.}~\bibnamefont
  {Deng}},\ and\ \bibinfo {author} {\bibfnamefont {X.}~\bibnamefont {Yang}},\
  }\bibfield  {title} {\bibinfo {title} {Large violation of the wiedemann-franz
  law driven by electron-hole compensation in the topological semimetal
  {CoSi}},\ }\href {https://doi.org/10.1103/5svq-g1k7} {\bibfield  {journal}
  {\bibinfo  {journal} {Phys. Rev. B}\ }\textbf {\bibinfo {volume} {113}},\
  \bibinfo {pages} {L121102} (\bibinfo {year} {2026})}\BibitemShut {NoStop}%
\bibitem [{\citenamefont {Weng}\ \emph {et~al.}(2016)\citenamefont {Weng},
  \citenamefont {Fang}, \citenamefont {Fang},\ and\ \citenamefont
  {Dai}}]{PhysRevB.93.241202}%
  \BibitemOpen
  \bibfield  {author} {\bibinfo {author} {\bibfnamefont {H.}~\bibnamefont
  {Weng}}, \bibinfo {author} {\bibfnamefont {C.}~\bibnamefont {Fang}}, \bibinfo
  {author} {\bibfnamefont {Z.}~\bibnamefont {Fang}},\ and\ \bibinfo {author}
  {\bibfnamefont {X.}~\bibnamefont {Dai}},\ }\bibfield  {title} {\bibinfo
  {title} {Topological semimetals with triply degenerate nodal points in
  $\ensuremath{\theta}$-phase tantalum nitride},\ }\href
  {https://doi.org/10.1103/PhysRevB.93.241202} {\bibfield  {journal} {\bibinfo
  {journal} {Phys. Rev. B}\ }\textbf {\bibinfo {volume} {93}},\ \bibinfo
  {pages} {241202} (\bibinfo {year} {2016})}\BibitemShut {NoStop}%
\bibitem [{\citenamefont {Rao}\ \emph {et~al.}(2019)\citenamefont {Rao},
  \citenamefont {Li}, \citenamefont {Zhang}, \citenamefont {Tian},
  \citenamefont {Li}, \citenamefont {Fu}, \citenamefont {Tang}, \citenamefont
  {Wang}, \citenamefont {Li}, \citenamefont {Fan} \emph
  {et~al.}}]{rao2019observation}%
  \BibitemOpen
  \bibfield  {author} {\bibinfo {author} {\bibfnamefont {Z.}~\bibnamefont
  {Rao}}, \bibinfo {author} {\bibfnamefont {H.}~\bibnamefont {Li}}, \bibinfo
  {author} {\bibfnamefont {T.}~\bibnamefont {Zhang}}, \bibinfo {author}
  {\bibfnamefont {S.}~\bibnamefont {Tian}}, \bibinfo {author} {\bibfnamefont
  {C.}~\bibnamefont {Li}}, \bibinfo {author} {\bibfnamefont {B.}~\bibnamefont
  {Fu}}, \bibinfo {author} {\bibfnamefont {C.}~\bibnamefont {Tang}}, \bibinfo
  {author} {\bibfnamefont {L.}~\bibnamefont {Wang}}, \bibinfo {author}
  {\bibfnamefont {Z.}~\bibnamefont {Li}}, \bibinfo {author} {\bibfnamefont
  {W.}~\bibnamefont {Fan}}, \emph {et~al.},\ }\bibfield  {title} {\bibinfo
  {title} {Observation of unconventional chiral fermions with long fermi arcs
  in {CoSi}},\ }\href {https://doi.org/10.1038/s41586-019-1031-8} {\bibfield
  {journal} {\bibinfo  {journal} {Nature}\ }\textbf {\bibinfo {volume} {567}},\
  \bibinfo {pages} {496} (\bibinfo {year} {2019})}\BibitemShut {NoStop}%
\bibitem [{Dat()}]{Dataava}%
  \BibitemOpen
  \bibfield  {title} {\bibinfo {title} {See the computational details at
  \url{git@github.com:CTCMP/dataset_papers.git}},\ }\href
  {git@github.com:CTCMP/dataset_papers.git} {\bibinfo  {journal}
  {arXiv2601.05522.zip}\ }\BibitemShut {NoStop}%
\end{thebibliography}
%

~~~\\
~~~\\
~~~\\

\end{document}